\documentclass[usenatbib]{mnras}

\usepackage{graphicx}
\usepackage{amssymb}
\usepackage{amsmath}
\usepackage{enumerate}
\usepackage{microtype}

\usepackage[usenames]{xcolor}
\hypersetup{colorlinks=true,citecolor=cyan}

\newcommand{\del}{\ensuremath{\partial}}
\newcommand{\Msun}{\ensuremath{M_{\odot}}}
\newcommand{\Mh}{\ensuremath{h^{-1}M_{\odot}}}
\newcommand{\Mpch}{\ensuremath{h^{-1}{\rm Mpc}}}

\newcommand{\de}{\ensuremath{{\rm d}}}

\newcommand{\eqn}[1]{equation~\eqref{#1}}

\newcommand{\fig}[1]{Figure~\ref{#1}}

\newcommand{\secn}[1]{Section~\ref{#1}}

\newcommand{\be}{\begin{equation}}
\newcommand{\ee}{\end{equation}}
\newcommand{\bear}{\begin{eqnarray}}
\newcommand{\ear}{\end{eqnarray}}
\newcommand{\f}{\frac}

\begin{document}
\pagerange{\pageref{firstpage}--\pageref{lastpage}} 

\title[Photon conservation and the 21 cm power spectrum]
{Photon number conservation and the large-scale 21 cm power spectrum in semi-numerical models of reionization}
\author[Choudhury \& Paranjape]
{T. Roy Choudhury$^1$\thanks{E-mail: tirth@ncra.tifr.res.in}
and
Aseem Paranjape$^2$\thanks{Email: aseem@iucaa.in}
\\
$^1$National Centre for Radio Astrophysics, TIFR, Post Bag 3, Ganeshkhind, Pune 411007, India\\
$^2$Inter-University Center for Astronomy \& Astrophysics, Post Bag 4, Ganeshkhind, Pune 411007, India
}

\maketitle
\label{firstpage}

\date{\today}

\begin{abstract}
Semi-numerical models of the reionization of neutral hydrogen (HI) based on the excursion set (ES) approach are known to violate photon number conservation at the few per cent level. 
In this work, we highlight a more severe, previously unrecognized shortcoming of ES models: the large-scale 21~cm power spectrum (equivalently, HI bias $b_{\rm HI}$) is a relatively strong function of the spatial resolution used to generate ES ionization maps. 
We trace this problem to the fact that photon non-conservation in these models arises from a resolution-dependent mixture of spatially resolved, photon non-conserving bubbles, and partially ionized grid cells which are perfectly photon-conserving by construction. 
We argue that this inevitably leads to a resolution-dependence of $b_{\rm HI}$, with the correct, converged value only emerging at very coarse resolution. 
Quantitatively, we find that $b_{\rm HI}$ can be non-converged by as much as $\sim20$-$25\%$ in conservative ES implementations with grid sizes $\Delta x = 5$-$10h^{-1}$cMpc, even when photon non-conservation is as small as $\sim3$-$4\%$. 
Thus, although numerically efficient, ES ionization maps coarse enough to produce a converged HI bias would wash out all topological features of the ionization field at scales $k\gtrsim0.05h/$cMpc. 
We therefore present a new, \emph{explicitly photon conserving} (PC) semi-numerical algorithm which distributes photons isotropically around sources while also accounting for anisotropic overlaps between nearby bubbles. 
Our PC algorithm predicts a \emph{resolution-independent} value of $b_{\rm HI}$ consistent with the result of low-resolution ES maps, thus serving as a useful compromise between standard ES implementations and more expensive radiative transfer simulations. 
\end{abstract}


\section{Introduction}
\label{outline}

In models where reionization of the cosmic neutral hydrogen (HI) is driven by sources in galaxies, the process is characterized by the growth and overlap of ionized ``bubbles'' \citep[for reviews see][]{2001PhR...349..125B,2006PhR...433..181F,2009CSci...97..841C}. A numerically inexpensive and reasonably accurate way of simulating the bubbles for such models is to use the excursion set (ES) formalism \citep{2007ApJ...669..663M,2007ApJ...654...12Z,2008MNRAS.386.1683G,2009MNRAS.394..960C,2010MNRAS.406.2421S,2011MNRAS.411..955M}. These semi-numerical models enable simulating the HI fluctuations in large volumes of reasonably high resolution, thus enabling computation of observable quantities relevant for the low-frequency radio telescopes (e.g., GMRT\footnote{\tt http://www.gmrt.ncra.tifr.res.in}, MWA\footnote{\tt https://www.haystack.mit.edu/ast/arrays/mwa}, LOFAR\footnote{\tt http://www.lofar.org}, PAPER\footnote{\tt http://eor.berkeley.edu}, HERA\footnote{\tt http://reionization.org}, SKA\footnote{\tt https://astronomers.skatelescope.org}). The use of these semi-numerical simulations is further justified as they are found to agree fairly well with the more accurate radiative transfer calculations \citep{2011MNRAS.414..727Z,2011MNRAS.411..955M,2014MNRAS.443.2843M}. In recent times, they are also found to be quite effective for parameter estimation \citep{2015MNRAS.449.4246G}.

In any model of reionization, the number of hydrogen atoms ionized must be equal to the number of ionizing photons produced by the sources (compensated for recombination). It turns out that the ES models violate this equality, i.e., they do not conserve the number of ionizing photons \citep{2005ApJ...630..643M,2007ApJ...654...12Z,2014MNRAS.442.1470P,2016MNRAS.460.1801P,2017MNRAS.468..122H}. Usually one gets around this difficulty by simply scaling the efficiency of the ionizing sources so as to produce the global average value of the neutral fraction as desired. For example, while comparing the semi-numerical models with the full radiative transfer calculations, this scaling is used to normalize the semi-numerical simulations to the same global neutral fraction as found from the radiative transfer simulations \citep{2011MNRAS.414..727Z,2014MNRAS.443.2843M}.

The cause of this non-conservation, at least in the analytical ES models, can be traced back to the fact that the ES approach keeps track of only average mass fractions instead of the stochastically fluctuating source counts. In our previous paper \citep{2016MNRAS.460.1801P}, we proposed a possible solution where the bubble growth is modelled using an (approximately) photon number conserving Monte Carlo approach by partitioning initial patches of dark matter into proto-haloes \citep{sl99b}. This scheme, however, suffered from the problem that it was strictly valid only for white noise initial conditions, with no straightforward generalization to cold dark matter power spectra.

In this paper, we focus our attention on the semi-numerical models now common in the literature. We examine in detail the magnitude and origin of photon non-conservation in these algorithms and explore its consequences. As we will show, even a small amount (few per cent) of photon non-conservation in ES-based semi-numerical models generically leads to a more severe problem, which is that the large-scale 21~cm power spectrum picks up a dependence on the spatial resolution at which ionization maps are generated. Given that these semi-numerical models are expected to play an important role in extracting reionization parameters from ongoing and upcoming 21~cm experiments, it is important to understand and address this problem, which is the goal of this work. 

Below, in addition to quantifying in detail the shortcomings of the ES approach as regards photon non-conservation and the non-convergence of the predicted 21~cm power spectrum, we will also present a new semi-numerical approach which is explicitly photon-conserving (PC) and which consequently produces a properly converged large-scale power spectrum. Our PC algorithm, which is numerically less efficient than standard ES implementations (while the ES algorithm completes under a second for typical resolutions considered in this work, the PC method takes $\sim$ hours, see \secn{sec:compare_PC_ES}), serves as a compromise between the ES models and the perfectly photon-conserving, but much more expensive, full radiative transfer simulations.

The plan of the paper is as follows: In \secn{sec:density_halo_fields}, we describe the dark matter and halo fields used in this work. In \secn{sec:photonNC}, we describe the variant of the ES approach we use as our baseline model, and explore in detail the nature and origin of photon non-conservation and power spectrum non-convergence at fixed redshift in this model. In \secn{sec:PC}, we present our new PC algorithm and compare its results to the ES approach, again at fixed redshift. In \secn{sec:realistic} we present a comparison as a function of redshift using a simple but realistic reionisation history. We summarize and conclude in \secn{sec:conclude}. The Appendices present details of some of the results used in the main text, as well as results using a publicly available variant of the ES approach.
Throughout, we use a flat $\Lambda$CDM cosmology with parameters given by $\Omega_m = 0.308, \Omega_b = 0.0482, h = 0.678, n_s = 0.961, \sigma_8 = 0.829$ \citep{2014A&A...571A..16P}. We quote the values of sizes and distances in $h^{-1}$cMpc while values of (halo) masses are quoted in $\Msun$.

\section{The density field and haloes}
\label{sec:density_halo_fields}

The very first ingredients of any reionization simulation are the underlying baryonic (mainly hydrogen) density field and the masses and locations of haloes that are capable of forming stars. For this purpose, it is often sufficient to generate the collisionless dark matter density field and assume that the baryonic fluctuations simply follow the dark matter fluctuations at scales of interest. For the analysis in this paper, we performed a dark matter-only $N$-body simulation using the publicly available code GADGET-2 \citep{2005MNRAS.364.1105S}\footnote{\tt https://wwwmpa.mpa-garching.mpg.de/gadget/}, with the initial conditions being generated by N-GenIC in a cubic periodic box of size $500 h^{-1}$cMpc with $512^3$ particles. The size of the box is chosen so as to probe the sufficiently large scales relevant for upcoming 21~cm probes like the SKA.

The identification of haloes is less straightforward because the particle mass of our simulation is $1.2 \times 10^{11} \Msun$ which is significantly larger than the mass of the smallest star-forming halo $\sim 10^8 \Msun$ (assuming only atomic cooling to be effective). As a result, it is not possible to identify the haloes using any group-finding algorithm.\footnote{Identifying haloes of mass as small as $10^8 \Msun$ in a cubical box of size $500 h^{-1}$cMpc would require running a $N$-body simulation with $\approx 12000^3$ particles, assuming a minimum of only 10 particles to characterize a halo.} In this case, one has to implement some sub-grid algorithm to find the low-mass haloes in the simulation box. In this work we use a semi-analytical prescription based on sampling a conditional mass function given by the fitting form provided by \citet{2002MNRAS.329...61S}. We describe the details of our prescription in Appendix~\ref{app:halofield}. 

The end result of this sub-grid sampling algorithm is the ``collapse fraction'' $f_{\rm coll}(\mathbf{x})$, which gives the fraction of mass in the cell at position $\mathbf{x}$ contained in haloes of mass larger than some threshold mass $M_{\rm min}$, and is the fundamental quantity used in simulating the ionization maps. This collapse fraction is generated on a grid with some resolution $\Delta x$, using the non-linear dark matter density contrast $\delta_{\rm NL}(\mathbf{x})$ as an input. 

For reasons discussed in Appendix~\ref{app:halofield}, in the main analysis of this work we will always generate the collapse fraction field using a grid resolution $\Delta x=5h^{-1}$cMpc, even when we want to study ionization maps generated at coarser resolutions. In spirit, our method is similar to running an $N$-body simulation at the best possible resolution, identifying all the haloes of interest in the box, and then smoothing both the density and halo fields to the desired resolution for generating the maps. Our method thus ensures that the resolution-dependent effects we discuss later arise only because of the algorithm used for generating the ionization maps, while the density and halo fields by construction converge to the same value at scales larger than the coarsest resolution under consideration.

The main observable in the ongoing and upcoming interferometric 21~cm experiments is the differential brightness temperature
\be
\delta T_b(\mathbf{x}) = \bar{T}_b~x_{\rm HI}(\mathbf{x}) [1 + \delta_{\rm NL}(\mathbf{x})],
\ee
where $x_{\rm HI}(\mathbf{x})$ is the neutral hydrogen fraction and $\bar{T}_b \approx 22 \mbox{mK} [(1 + z) / 7]^{1/2}$.  We assume here that the 21~cm spin temperature is significantly higher than the CMB temperature and hence does not contribute to the $\delta T_b$ fluctuations. Note that the global average of $\delta T_b$ is given by
\be
\langle \delta T_b(\mathbf{x}) \rangle = \bar{T}_b (1 -  Q_{\rm HII}^M).
\ee
For convenience we prefer to work with the dimensionless quantity
\be
\Delta_{\rm HI}(\mathbf{x}) \equiv \f{\delta T_b(\mathbf{x})}{\langle \delta T_b(\mathbf{x}) \rangle} = \f{x_{\rm HI}(\mathbf{x}) [1 + \delta_{\rm NL}(\mathbf{x})]}{1 - Q_{\rm HII}^M},
\ee
which measures the fluctuations in the HI density. Defined like this, $\langle \Delta_{\rm HI}(\mathbf{x}) \rangle = 1$ by construction. We denote the power spectrum of $\Delta_{\rm HI}$ by $P_{\rm HI}(k)$ (which is isotropic since we do not include the effect of peculiar velocities and other line of sight effects in this work). A relevant quantity is the HI bias $b_{\rm HI}(k)$ defined as
\be
b_{\rm HI}^2(k) \equiv \f{P_{\rm HI}(k)}{P(k)},
\label{eq:b_HI_k}
\ee
where $P(k)$ is the matter power spectrum at the redshift of interest. While calculating $b_{\rm HI}^2(k)$ below, we compute $P(k)$ from the simulation box itself so as to minimize the effects of sample variance at large scales.

\section{Photon non-conservation and its consequences}
\label{sec:photonNC}
In this section we discuss in detail the issue of photon non-conservation in excursion set (ES) models of reionization and demonstrate that it leads to a (generically more severe) problem of non-convergence of the large-scale bias $b_{\rm HI}(k)$. We start with a description of the ES method itself.

\subsection{Semi-numerical excursion set (ES) model of reionization}
The ES method of generating ionized regions during reionization is based on identifying spherical regions that can be ``self-ionized'' \citep{2004ApJ...613....1F}. The essential summary of the method is given below, for details we refer the reader to \citet{2007ApJ...654...12Z,2007ApJ...669..663M,2009MNRAS.394..960C,2010MNRAS.406.2421S,2014MNRAS.443.2843M}. Our implementation of this method is representative of most variations found in the literature; however, for completeness we have also performed some key aspects of our analysis using the publicly available 21cmFAST code\footnote{\tt https://github.com/andreimesinger/21cmFAST} \citep{2011MNRAS.411..955M}.

We assume that the haloes have ionizing emissivities proportional to their (dark matter) mass. Given the value of $f_{\rm coll}(\mathbf{x})$ and $\delta_{\rm NL}(\mathbf{x})$ in grid cells of the simulation box, a location $\mathbf{x}$ is flagged as ionized if, within a spherical region of radius $R$ around it, the condition
\be
\zeta f_{\rm coll}(\mathbf{x},R) \geq 1,
\label{eq:barrier_condition}
\ee
is satisfied for any value of $R$, where $f_{\rm coll}(\mathbf{x},R)$ is the collapsed mass fraction within the spherical volume. The parameter $\zeta$ is the ``effective'' ionizing efficiency, representing the number of photons in the IGM per hydrogen atom in stars, compensated for the number of hydrogen recombinations in the IGM (assuming it to be uniform). Note that, unlike the original ES approach for haloes, which operates in the initial conditions, the ES-based semi-numerical models for reionization work with the \emph{non-linear} dark matter density field, so that $f_{\rm coll}(\mathbf{x},R) = \langle f_{\rm coll}(\mathbf{x})~(1 + \delta_{\rm NL}(\mathbf{x}) \rangle_R$, where $\langle \ldots \rangle_R$ denotes the average over a spherical volume of radius $R$. As $R \to \infty$, the quantity $f_{\rm coll}(\mathbf{x},R)$ approaches the value of the global mean collapse fraction. 
In this work, we discuss models employing two different filters for calculating $f_{\rm coll}(\mathbf{x},R)$, namely the spherical tophat filter in real space and a filter that is tophat in the $k$-space (commonly called the sharp-$k$ filter). 

Points which do not satisfy condition (\ref{eq:barrier_condition}) are assigned an ionized fraction $x_{\rm HII}(\mathbf{x}) = \zeta f_{\rm coll}(\mathbf{x})$, with $f_{\rm coll}(\mathbf{x})$ being the collapse fraction calculated at the resolution of the grid which is used for generating the ionization field. We will refer to these points, which will play an important role in understanding the results below, as ``partially ionized'' cells. 

In the following, we explore ionization maps generated at different resolutions (i.e., different grid sizes $\Delta x$). As mentioned earlier, the finest resolution we work with corresponds to $\Delta x = 5 h^{-1}$cMpc (for the box of size $500 h^{-1}$cMpc). Irrespective of the resolution of the ionization map, we always begin with the density field $\delta_{\rm NL}(\mathbf{x})$ and the collapse fraction $f_{\rm coll}(\mathbf{x})$ generated at $\Delta x = 5 h^{-1}$cMpc. For the case where we desire to generate an ionization map at a coarser resolution $\Delta x > 5 h^{-1}$cMpc, we smooth both the fields to the desired resolution using a boxcar filter. The ES formalism for generating the maps is then applied on these smoothed fields.

Given the method of generating the ionization field above, it is clear the number of ionizing photons (adjusted for recombinations) produced by the sources in haloes is $\zeta f_{\rm coll}$, where $f_{\rm coll}$ is the global mean collapse fraction. It can be shown from simple theoretical arguments \citep{2016MNRAS.460.1801P} that the number of photons produced $\zeta f_{\rm coll}$ would be identically equal to the mass-averaged (or the Lagrangian) ionized fraction
\be
Q_{\rm HII}^M \equiv \langle x_{\rm HII}(\mathbf{x})~[1 + \delta_{\rm NL}(\mathbf{x})] \rangle = \langle [1 - x_{\rm HI}(\mathbf{x})]~[1 + \delta_{\rm NL}(\mathbf{x})] \rangle.
\ee
Any deviation of the ratio $\zeta f_{\rm coll} / Q_{\rm HII}^M$ from unity would imply that photon number is \emph{not} being conserved.

\begin{figure}
\includegraphics[width=0.5\textwidth]{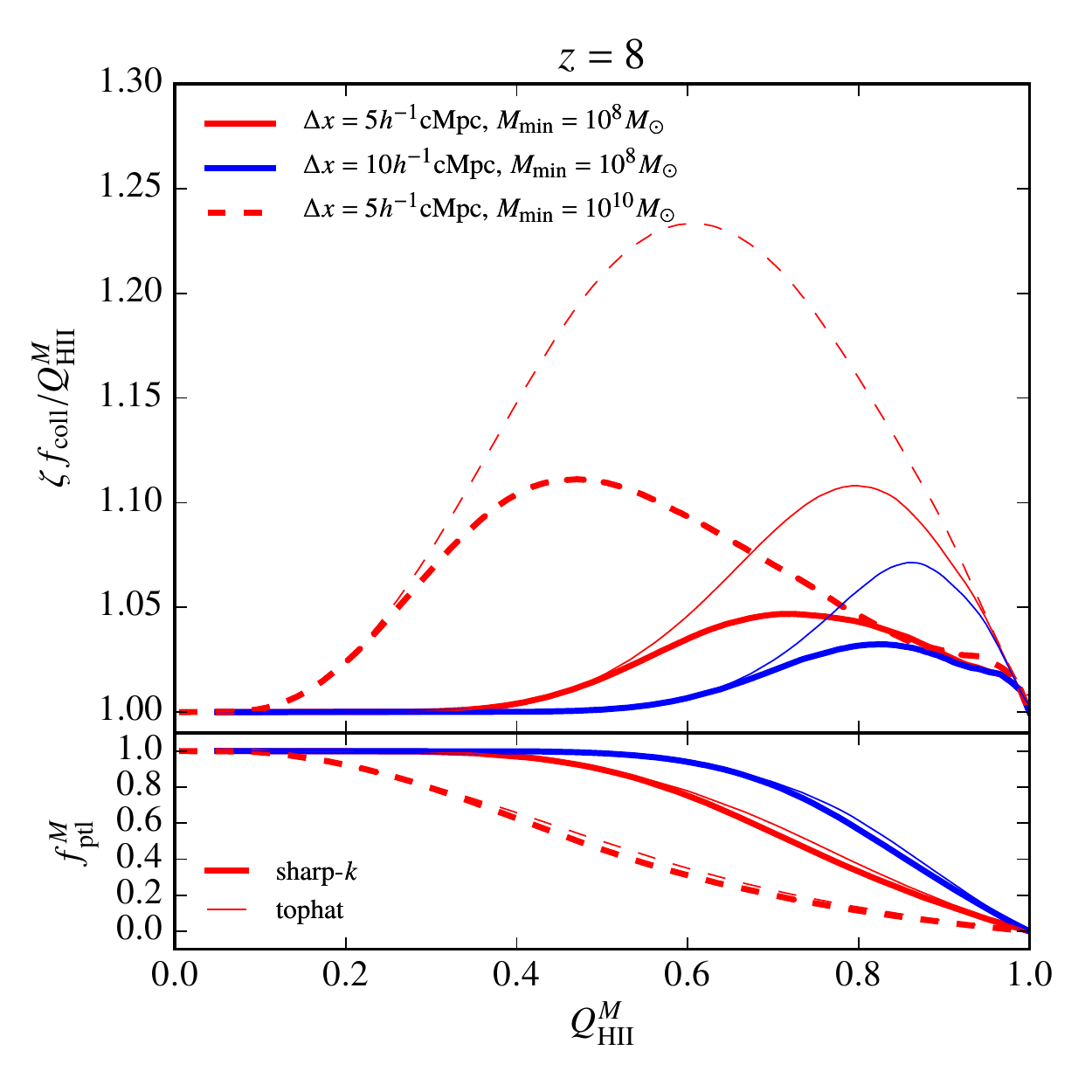}
\caption{
\emph{Top panel:} The ratio $\zeta f_{\rm coll} / Q^M_{\rm HII}$ as a function of $Q^M_{\rm HII}$ for $z = 8$ for a neutral hydrogen field obtained using a conventional excursion set formalism. Different values of $Q^M_{\rm HII}$ in the plot are obtained by varying the value of $\zeta$ keeping the underlying density same. The thick lines are obtained using the sharp-$k$ filter while the thin lines are for the spherical tophat filter. The parameter values are indicated in the legend. \emph{Bottom panel:} The fraction $f_{\rm ptl}^M$ of ionized mass in partially ionized cells for the same parameter values. 
}
\label{fig:zetafcoll_by_Q_fixed_z}
\end{figure}

\begin{figure*}
\includegraphics[width=\textwidth]{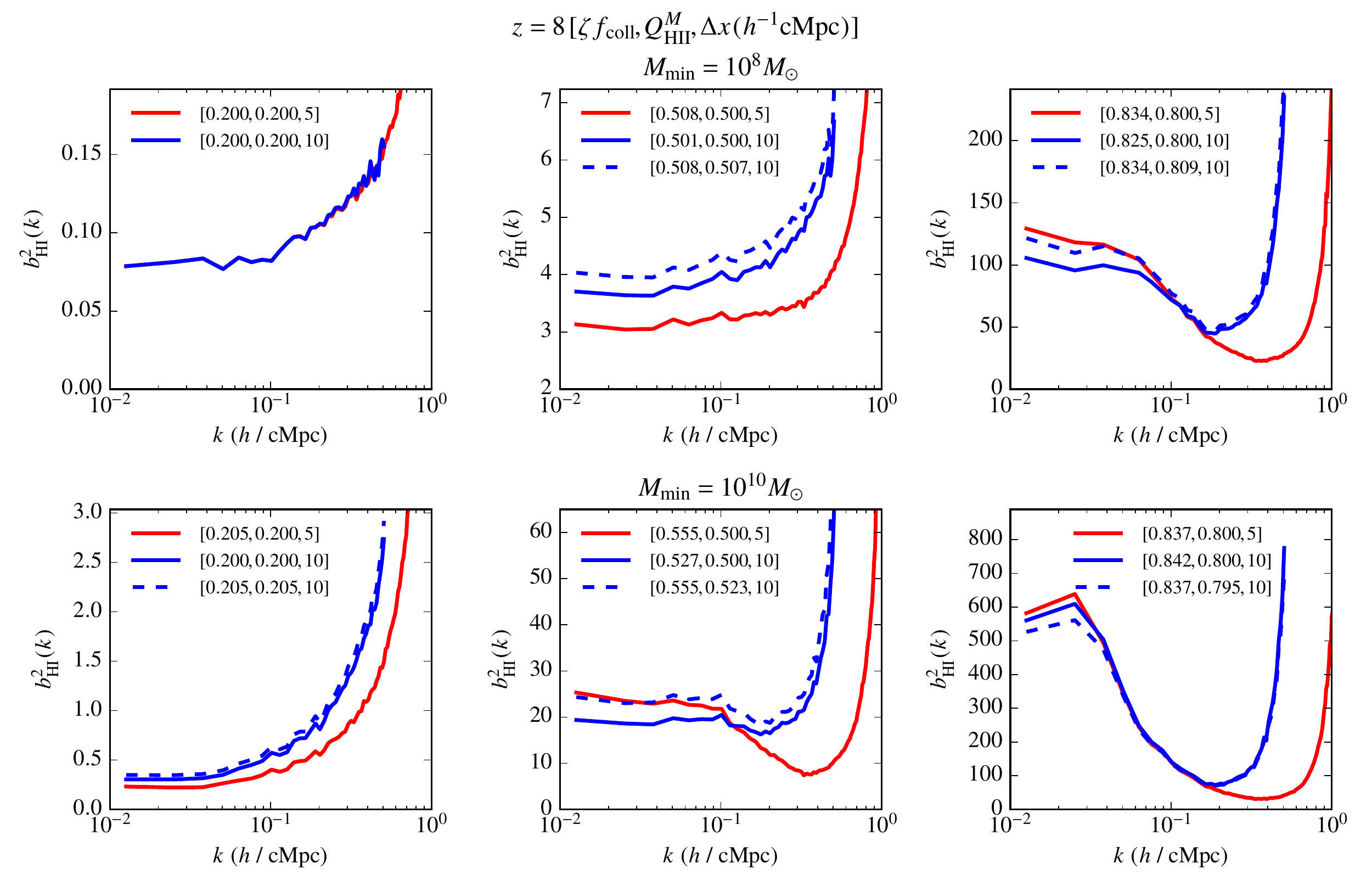}
\caption{
The bias $b_{\rm HI}^2(k) = P_{\rm HI}(k) / P(k)$ of the HI density fluctuations for the ES models using the sharp-$k$ filter. Different panels and curves are for different resolution and parameter values as indicated in the respective legends.
}
\label{fig:bias_6panels_onlyES}
\end{figure*}

\subsection{Photon non-conservation in ES models}
\label{subsec:ES-photonNC}
In order to study photon (non-)conservation, we first consider the case where the density and the halo fields are fixed to a particular redshift, say, $z = 8$. We vary the value of $\zeta$ to obtain different values of $Q_{\rm HII}^M$ and hence study the IGM at different stages of ionization. This allows us to disentangle any effect of the evolution of the underlying density field from the ionization maps. 

The top panel of \fig{fig:zetafcoll_by_Q_fixed_z} shows the measure of photon non-conservation, $\zeta f_{\rm coll} / Q_{\rm HII}^M$, plotted against $Q_{\rm HII}^M$ for different values of $\Delta x$ and $M_{\rm min}$ and for different filters. The value of $M_{\rm min} = 10^8 \Msun$ corresponds to the case where only atomically cooled haloes contribute to ionizing photons, while $M_{\rm min} = 10^{10} \Msun$ would correspond to cases where, e.g., the photons are unable to escape from the low-mass galaxies \citep[see, e.g.,][]{2008ApJ...672..765G} or the reionization is driven by quasar-like sources residing in high-mass haloes \citep{2017MNRAS.469.4283K}. The bottom panel of the Figure shows the fraction $f_{\rm ptl}^M$ of ionized mass that are contained in partially ionized cells (i.e., the cells where the barrier crossing condition (\ref{eq:barrier_condition}) is not satisfied for any value of $R$) in the simulation box.

We see that all the cases conserve photons in the early stages of reionization. This is because the sizes of the ionized bubbles are smaller than the grid resolution and hence almost all the cells are partially ionized (as can be seen from the lower panel of the Figure). These cells, by construction, conserve photons as they are assigned an ionized fraction $x_{\rm HII}(\mathbf{x}) = \zeta f_{\rm coll}(\mathbf{x})$. For a fixed resolution, the conservation holds for larger values of $Q_{\rm HII}^M$ at smaller $M_{\rm min}$ as the characteristic bubble sizes are smaller. On the other extreme, the models approach photon conservation towards the very late stages $Q_{\rm HII}^M \to 1$ where the ionization regions are expected to grow to sizes comparable to the simulation volume and the fluctuations in the ionization maps are caused by the neutral ``islands'' far away from the ionization sources. At this stage, the details of the ES method for generating ionized bubbles become less critical. 
The non-conservation is most prominent in the intermediate stages which involve a mixture of partially ionized cells and fully resolved bubbles identified as collections of neighbouring fully ionized cells. The key point which will be relevant below is that, \emph{for fixed $\zeta$ and $M_{\rm min}$, the amount of non-conservation is resolution-dependent}.

Interestingly, we find that the non-conservation is less severe for the maps made with the sharp-$k$ filter than those made using the spherical tophat filter. Hence for the rest of the paper, we would present our results only for the sharp-$k$ filter keeping in mind that all the conclusions related to the shortcomings of the ES-based models would be stronger for the tophat filter.

\subsection{Photon non-conservation implies bias non-convergence}
\label{subsec:photonNC->biasNC}
The amount of non-conservation of photon numbers in a generic ES model depends on various factors, e.g., the model parameters ($\zeta$ and $M_{\rm min}$), the filter used for identifying ionized regions, and also, as we emphasized, the resolution used to generate the maps. One might argue that, for a carefully chosen filter (say, sharp-$k$) and realistic model parameters, the non-conservation is never too large (within $\sim 5\%$), particularly if one decides to work at a low resolution (say, $\Delta x \gtrsim 5 h^{-1}$cMpc). In that case one expects an error of a few per cent in the globally averaged ionized mass fraction, which is probably acceptable given that much of the physics at high redshift is poorly understood. 
As we argue next, however, \emph{the resolution-dependence of photon non-conservation also generically results in a resolution-dependence of the predicted large-scale 21~cm power spectrum}.\footnote{For this discussion, we will assume that $M_{\rm min}$ is always fixed to a single value.}

This is easiest to anticipate when working at fixed ionized mass fraction $Q_{\rm HII}^M$. Consider generating maps at two resolutions $\Delta x_1$ and $\Delta x_2$, for some choice of parameters $\zeta$ and $M_{\rm min}$. In the standard approach followed in the literature, working at fixed $Q_{\rm HII}^M$ means that the value of $\zeta$ in each case is adjusted (to $\zeta=\zeta_1$ and $\zeta=\zeta_2$, say) such that the resulting value of $Q_{\rm HII}^M$ in the two maps is the same (and matched, e.g., to a radiative transfer simulation, or to any fixed value). The key point to note is that, since the level of photon non-conservation depends on resolution, $\zeta_1\neq\zeta_2$ in general. Since $\zeta$ controls the characteristic bubble sizes (or the bubble mass function), this immediately tells us that the large-scale clustering of the bubbles, and therefore of the 21~cm signal, will be different in the two maps. Thus, the prediction for large-scale bias picks up a resolution-dependence at fixed $Q_{\rm HII}^M$.

In fact, one can argue that there will also be a resolution-dependence of bias at fixed ionizing efficiency $\zeta$, although for a different reason. In this case, consider first the situation at very coarse resolution, such that essentially no bubble is resolved in the maps. In this case, photon numbers are perfectly conserved because all cells are partially ionized. Moreover, as we show in Appendix~\ref{app:bias_ES}, at sufficiently large scales, the HI bias in this case is completely determined by the large-scale \emph{halo} bias which is perfectly converged by construction. 
Now consider the opposite extreme of a map generated at very high resolution, such that \emph{all} bubbles are resolved. 
This map will have a substantially higher level of photon non-conservation than the low-resolution one \citep[since the resolved ES bubbles do not conserve photon numbers, see][]{2016MNRAS.460.1801P}, and the clustering properties of the (unique) ionization field produced by the algorithm will therefore be different, in general, from the correct large-scale answer obtained at low resolution. 
So one goes from the converged, correct large-scale bias at very low resolution to another converged, but incorrect bias at very high resolution, passing through a non-converged set of values at intermediate resolutions. We show this explicitly in \fig{fig:unnormalized_bias_ionized_4panels_haloes} discussed in Appendix~\ref{app:bias_ES}. Clearly, the source of this resolution-dependence is again the fact that the level of photon non-conservation in the algorithm is resolution-dependent and leads to different maps at low and high resolution.

\begin{figure*}
\includegraphics[width=0.6\textwidth]{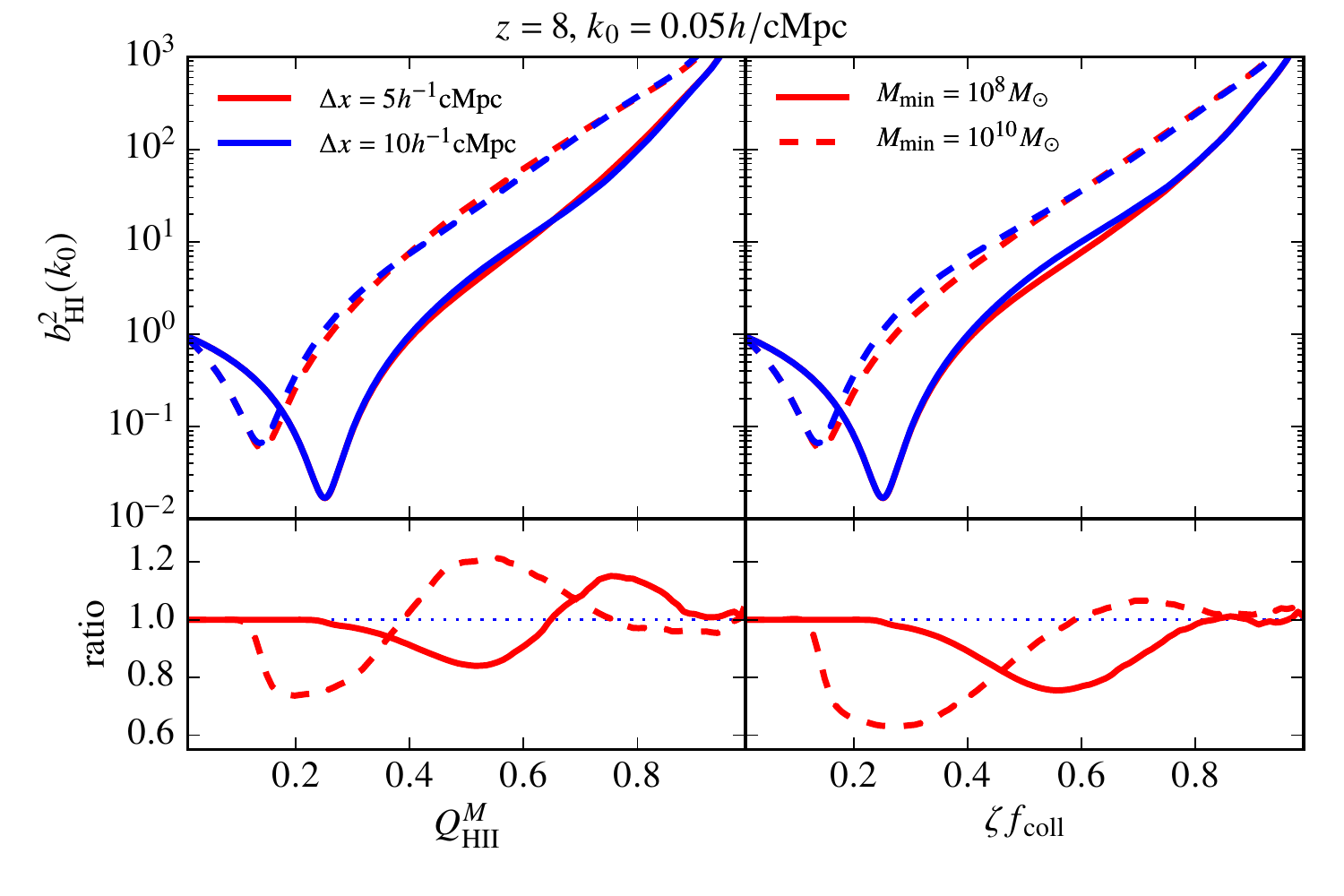}
\caption{
\emph{Top panels:} $b_{\rm HI}^2(k_0)$ for a representative $k = k_0 = 0.05~h/$cMpc as a function of $Q_{\rm HII}^M$ (left panel) and $\zeta f_{\rm coll}$ (right panel) for the ES model with sharp-$k$ filter. The different values of the parameters are indicated in the legend. \emph{Bottom panels:} The ratio $\delta b_{\rm HI}^2(k_0)$ of $b_{\rm HI}^2(k_0)$ for $\Delta x = 5 h^{-1}$cMpc to that for $\Delta x = 10 h^{-1}$cMpc for the same parameter values.
}
\label{fig:Pk_fixed_k_fixed_z}
\end{figure*}

The arguments above only indicate that there must be some level of bias non-convergence as a function of map resolution in the ES method. In the next section, we quantify the magnitude of this effect by measuring the 21~cm power spectrum in our simulated maps.

\subsection{Quantifying bias non-convergence in ES models}
\fig{fig:bias_6panels_onlyES} shows $b_{\rm HI}^2(k)$, defined in \eqn{eq:b_HI_k}, for different model parameters.
The top (bottom) panels are for the case $M_{\rm min} = 10^8 (10^{10}) \Msun$. The left, middle and right panels are for three representative values of the ionized mass fraction, namely, $Q_{\rm HII}^M = 0.2, 0.5, 0.8$ respectively (for the default resolution map). The legends in the respective panels also indicate the values of $\zeta f_{\rm coll}$ which are useful for determining the amount of photon non-conservation. In each panel, the red line corresponds to the map generated in the default resolution $\Delta x = 5 h^{-1}$cMpc. The corresponding blue lines are for the maps generated at a coarser resolution $\Delta x = 10 h^{-1}$cMpc. The solid blue line corresponds to the case where the value of $Q_{\rm HII}^M$ is chosen to be the same as the default resolution map, while the dashed blue line has the same value of $\zeta f_{\rm coll}$ as the default resolution map. All the relevant parameter values can be read off from legends as well.

The main point to note from the Figure is that, in general, the large-scale bias for the low-resolution maps \emph{does not converge} to that for the default resolution. As anticipated above, this non-convergence exists irrespective of whether the maps are normalized to the same of $Q_{\rm HII}^M$ or to the same $\zeta f_{\rm coll}$. The non-convergence is relatively less severe for small $Q_{\rm HII}^M$ (in fact it is non-existent when the maps are photon-conserving), but can be significantly large for late stages of reionization. 

To study how this non-convergence evolves as the reionization progresses, we plot $b_{\rm HI}^2(k_0)$ for a representative value of $k = k_0 = 0.05 h$/cMpc as a function of $Q^M_{\rm HII}$ (top left panel) and $\zeta f_{\rm coll}$ (top right panel) of \fig{fig:Pk_fixed_k_fixed_z}. This value of $k_0$ corresponds to scales that are significantly larger than either of the resolutions considered and is typical of scales to be observed by the first generation of radio telescopes \citep[see, e.g.,][]{2017ApJ...838...65P}. Our main results would remain qualitatively unchanged for any $k \lesssim 0.1 h$/cMpc. Various curves are for different resolutions and $M_{\rm min}$, as indicated in the legends. In the bottom panels of the same Figure, we plot the ratio of the bias for the two different resolutions, $[b_{\rm HI}^2(k_0)]_{\Delta x = 5 h^{-1}~{\rm cMpc}} / [b_{\rm HI}^2(k_0)]_{\Delta x = 10 h^{-1}~{\rm cMpc}}$. 

As one can see, the large-scale bias for the two different resolutions converges for smaller $Q_{\rm HII}^M$ only when the photon conservation holds (compare with \fig{fig:zetafcoll_by_Q_fixed_z}). In the stages where photons are not conserved, we find that the large-scale bias depends on the map resolution and \emph{the non-convergence between the two resolutions can be as high as $\sim 15$-$20\%$ at fixed $Q_{\rm HII}^M$, even though the photon non-conservation is $\lesssim 5\%$ for these models}. Another point to note is that, since the bias has \emph{not} converged, a proper estimate of the response of bias non-convergence to photon non-conservation would require the correct, converged value of bias at large scales. We will return to this point later. 

It is possible to understand the dependence of the large-scale bias  on various parameters in the ES-based models, as well as the complicated-looking behaviour as a function of $Q_{\rm HII}^M$ or $\zeta f_{\rm coll}$, using the analytical formalism of \citet{2004ApJ...613....1F}. We present a detailed discussion on this in Appendix \ref{app:bias_ES} and keep the focus of the main text on the issue of photon non-conservation and bias non-convergence. Unsurprisingly,  the bias non-convergence exists in other implementations of the ES-based semi-numerical simulations too, e.g., the publicly available 21cmFAST \citep{2011MNRAS.411..955M}; this is illustrated in Appendix \ref{app:bias_21cmFAST}.

\section{An explicitly photon-conserving model of reionization}
\label{sec:PC}

Given the results of the previous Section, there is a clear need for an explicitly photon-conserving (PC) model of reionization, unless one is willing to give up completely on resolving any details of the topology of the ionized volume during reionization. We will take the point of view that details of this topology are, in fact, interesting enough that it is worth investing in PC models. Below, we present one such implementation, without worrying about the relative efficiency of our method as compared to the ES prescriptions. Our proposed algorithm is thus numerically much slower than the ES-based models; however, it is still sufficiently efficient and flexible (compared to, say, radiative transfer simulations) to be run for many different sets of parameters in a reasonable amount of time.

\subsection{Description of the photon-conserving model}

As is the case with any model of reionization, we begin with the density and collapse fraction fields at a given $z$ at an appropriate resolution (described in \secn{sec:density_halo_fields}). Given the value of $\zeta$, we know how many ionizing photons are being produced by each grid cell in the box. Note that, although each grid cell may contain numerous sources, we can effectively treat it as a single source producing some number of photons.

Our PC model is based on constructing approximately spherical ionized regions around sources. The algorithm consists of two main rounds:

\begin{itemize}

\item In the first round, we assign ionized spheres of appropriate sizes around the radiation sources. Given a grid cell with $\zeta f_{\rm coll}(\mathbf{x}) > 0$ (let us call it a ``source'' cell), we distribute the ionizing photons starting from the cell itself followed by other cells in increasing order of distance from the original cell (keeping track of the periodic boundary conditions in the simulation box while calculating the distance between the cells). 

To be more specific, we first calculate the number of photons $N_{\gamma}(\mathbf{x}) = \zeta f_{\rm coll}(\mathbf{x}) [1 + \delta_{\rm NL}(\mathbf{x})] \bar{n}_H$ originating from the sources within the cell. Out of these, we assume that $n_H(\mathbf{x}) = [1 + \delta_{\rm NL}(\mathbf{x})] \bar{n}_H$ number of photons would be consumed by the hydrogen atoms in the source cell itself. The excess photons are then distributed to the cells which are nearest to the original cell. In case the number of photons available is more than the number of hydrogen atoms in these cells, they are flagged as completely ionized and we are left with an excess of unabsorbed photons. 

\begin{figure*}
  \includegraphics[width=0.9\textwidth]{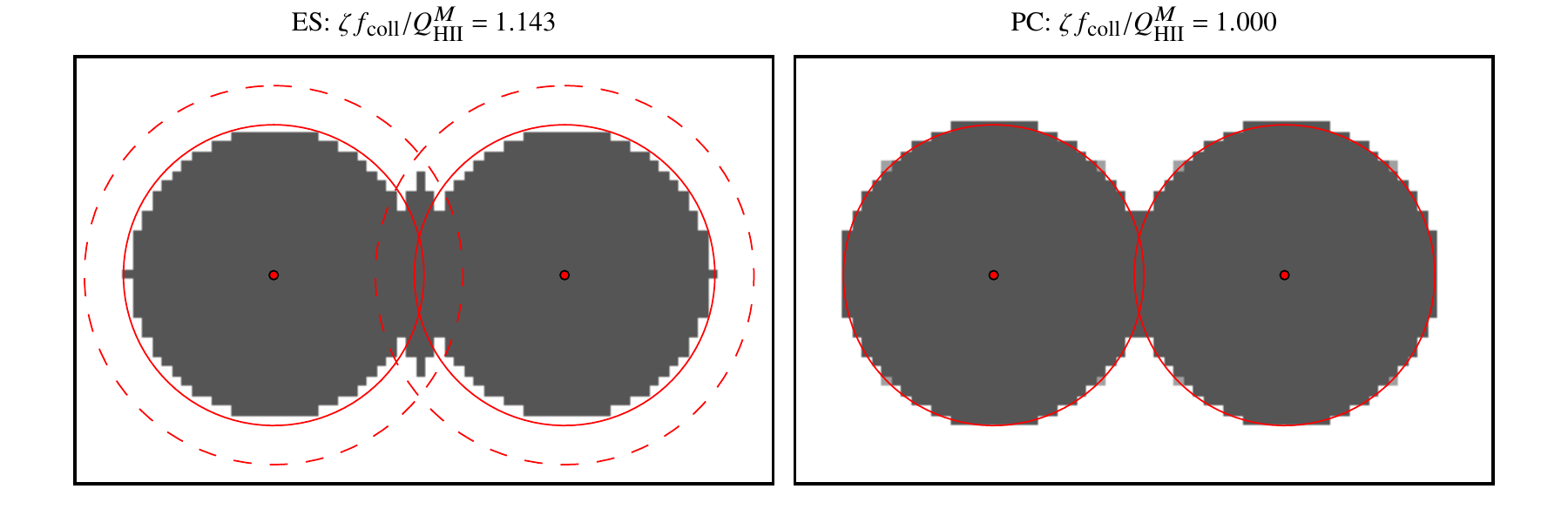}
  \caption{
  The projected ionization map for a toy model with two identical sources (indicated by the red dots) obtained using the excursion set model with spherical tophat filter \emph{(left panel)} and the photon-conserving model \emph{(right panel)}. The two-dimensional slice has been chosen to lie in the plane containing the two sources. The solid red circles indicate the ionized bubble size for the individual sources. In the left panel, the dashed red circles indicate the bubble size for sources of double the emissivity.
  }
  \label{fig:2source}
\end{figure*}

These excess photons are then distributed to the next nearest cells and the process continues until, for a set of cells at the same distance from the original source cells, we do not have enough photons to ionized all of them. In that case we simply distribute the available photons equally within these cells. Each of these cells is assigned an ionized fraction equal to the ratio of the number of photons available for that specific cell and the number of hydrogen atoms. At this point, we would have consumed all the photons that were produced in the original source cell. The entire process follows the physical intuition that individual sources emit photons isotropically.

We repeat the above process for all cells with $\zeta f_{\rm coll}(\mathbf{x}) > 0$. We carry out the process for each cell independently of the others, and hence any grid point which is ionized by more than one source cell can end up with an unphysical ionized fraction $x_{\rm HII}(\mathbf{x}) > 1$. These ``over-ionized'' cells, which arise because of overlapping ionized bubbles, are dealt with separately in the following round.

\item In the second round, the over-ionized cells with $x_{\rm HII}(\mathbf{x}) > 1$ from the first round are assumed to be effective sources with $N^{(2)}_{\gamma}(\mathbf{x}) = [x_{\rm HII}(\mathbf{x}) - 1] [1 + \delta_{\rm NL}(\mathbf{x})] \bar{n}_H$ number of photons. For such a cell, these photons are used to ionize other cells with the nearest cells being ionized first. While carrying out this process, we ensure that cells which already have $x_{\rm HII}(\mathbf{x}) \geq 1$ at the end of the first round are left unaffected, while cells which have  $x_{\rm HII}(\mathbf{x}) < 1$ at the end of the first round are allowed to consume photons (if available) required to completely ionize them. 

In case we are left with a situation where the number of hydrogen atoms in cells at a given distance (accounting for already ionized cells in the first round)  is larger than the number of photons available at that stage, we simply distribute the photons equally within the cells. Thus, individual sources are still assumed to be isotropic emitters. We then repeat the process for all over-ionized cells with $x_{\rm HII}(\mathbf{x}) > 1$, very similar to the first round.\footnote{Interestingly, despite the assumed locally isotropic nature of the emitting sources in each round, the \emph{global} ionization fronts would tend to be driven by relatively stronger sources in our method, consistent with physical expectations. Since the (over)ionized cells are not allowed to consume any excess photons in the second round, these are preferentially consumed in (partially) neutral cells which are likely to be nearer to the relatively weaker sources. Hence the apparent direction of the boundary of the ionized regions would be away from the stronger sources and towards the weaker sources.}

One crucial difference compared to the first round is that the process for a given over-ionized cell is \emph{not} independent of the others. While carrying out the process from the second over-ionized cell, we keep track of the ionization structure produced by the earlier over-ionized cells (e.g., a cell which is completely ionized by photons from an earlier over-ionized cell will not be affected by subsequent over-ionized cells). This helps ensure that we are able to deal with the overlapping regions at one go. Strictly speaking, this also introduces a dependence of the final ionization maps on the order in which over-ionized regions are dealt with, i.e., the resulting maps are not unique. Using minor modifications of the algorithm, we have verified that, in practice, this non-uniqueness is actually restricted to length scales not much larger than the grid scale. This will also be evident below when we show that the large-scale 21~cm power spectrum produced by our PC algorithm is manifestly independent of spatial resolution.

\end{itemize}

Overall, the method outlined above, which first allows cells to become over-ionized and then redistributes their photons to partially ionized neighbours, gels well with the physical intuition that the topology of ionized regions should respond to the anisotropies in the distribution of the IGM, despite individual emitters being isotropic. Since we explicitly track all the photons produced by sources, \emph{our method conserves photon number by construction}. 

This method is very similar to that used to treat the overlap of ionized bubbles in the one-dimensional radiative transfer simulations of \citet{2015MNRAS.447.1806G,2015MNRAS.453.3143G}, which has been found to agree well with the full radiative transfer simulations \citep{2018MNRAS.476.1741G}. An alternate way to treat the overlaps, where one increases the sizes of the overlapping spheres appropriately, has been implemented in the simulations of \citet{2009MNRAS.393...32T,2011MNRAS.410.1377T}, however, that method is somewhat slower than ours. Similar to the ES methods, our algorithm ensures that the bubbles are centred around the sources, thus maintaining the inside-out nature of reionization. A subtle difference between the ES and PC methods lies in the assignment of the partially ionized grid cells: while the ES method always assigns a ionized fraction $\zeta f_{\rm coll}$ to such cells, the ionized fraction for a partially ionized cell in our PC method can be larger than $\zeta f_{\rm coll}$ in that cell.

\subsection{Comparison between the photon-conserving and excursion set models}
\label{sec:compare_PC_ES}

Let us first compare our PC method with the ES method for a ``toy'' situation where we consider two identical sources in a medium of uniform density. This situation has been studied for understanding photon non-conservation by \citet{2007ApJ...654...12Z}. The results are shown in \fig{fig:2source}. The left panel \citep[which reproduces Figure 11 of][]{2007ApJ...654...12Z} is the result for the ES model with a spherical tophat filter.\footnote{Using a sharp-$k$ filter for this idealized model leads to artificial ``ringing'' features in the map, hence we show the results only for the spherical tophat filter.} The emissivity of the sources has been chosen so that their individual ionized bubbles (indicated by red solid lines) have non-zero overlap. One can see that the ES model produces an unphysical feature in the overlapping region between the spheres of radius $r_1$ and $2^{1/3} r_1$ (dashed red lines), where $r_1$ is the radii of the individual bubbles. The photon non-conservation in this case is $\sim 14\%$.

\begin{figure*}
  \includegraphics[width=1.1\textwidth]{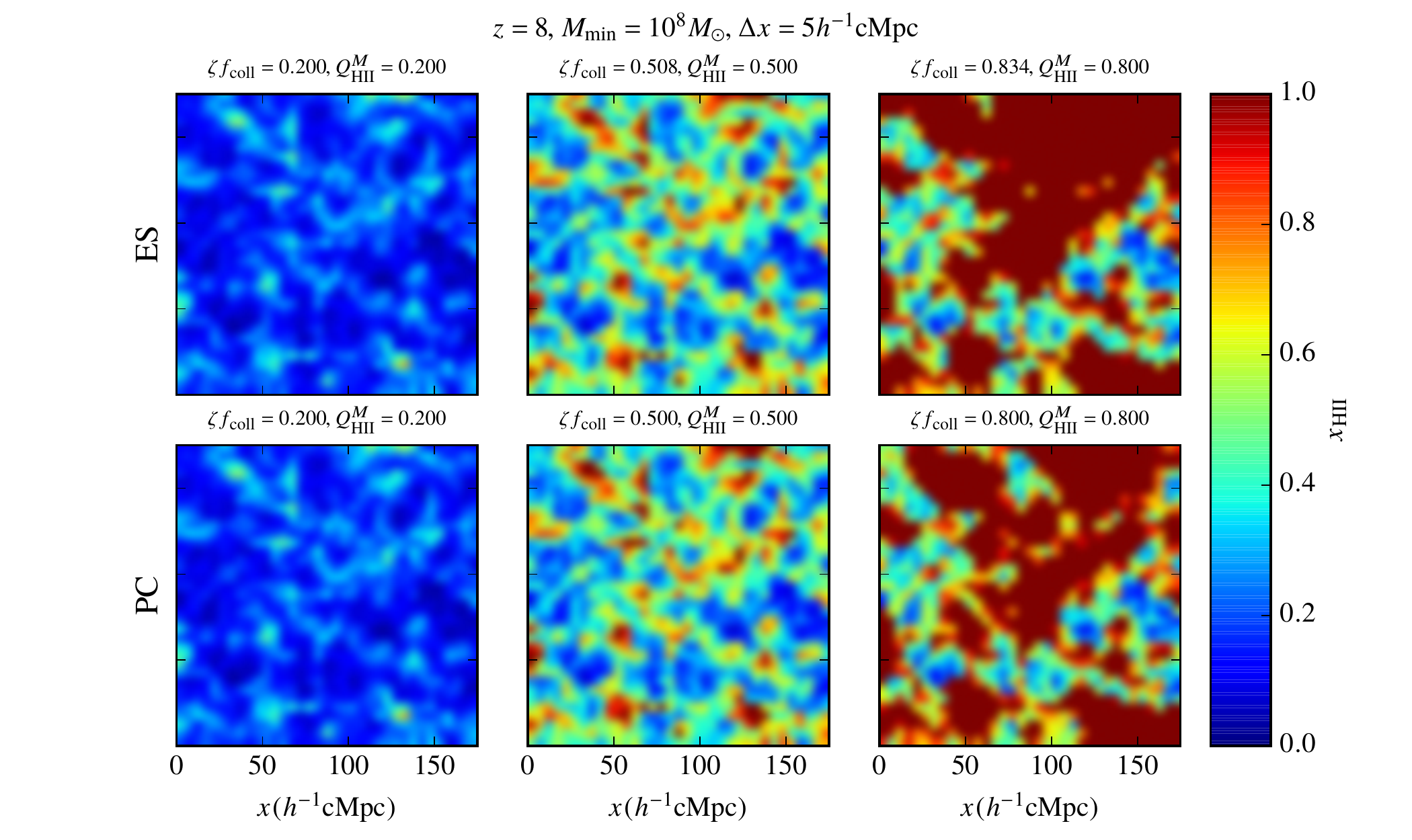}
  \caption{
  The two-dimensional slices of the ionized fraction $x_{\rm HII}$ obtained using the excursion set model with sharp-$k$ filter \emph{(top panels)} and the photon-conserving model \emph{(bottom panels)} at $z = 8$. The values of the parameters are indicated at the top of each panel.
  }
  \label{fig:plotionizedmaps}
\end{figure*}

The corresponding result from our PC scheme is shown in the right panel. We can see that the scheme does not give rise to any unphysical regions. Instead it distributes the excess photons in the overlapping region to the adjacent grid cells. Thus our PC scheme produces ionization maps which are in line with general physical intuitions for the simplest toy scenario.

\begin{figure*}
\includegraphics[width=\textwidth]{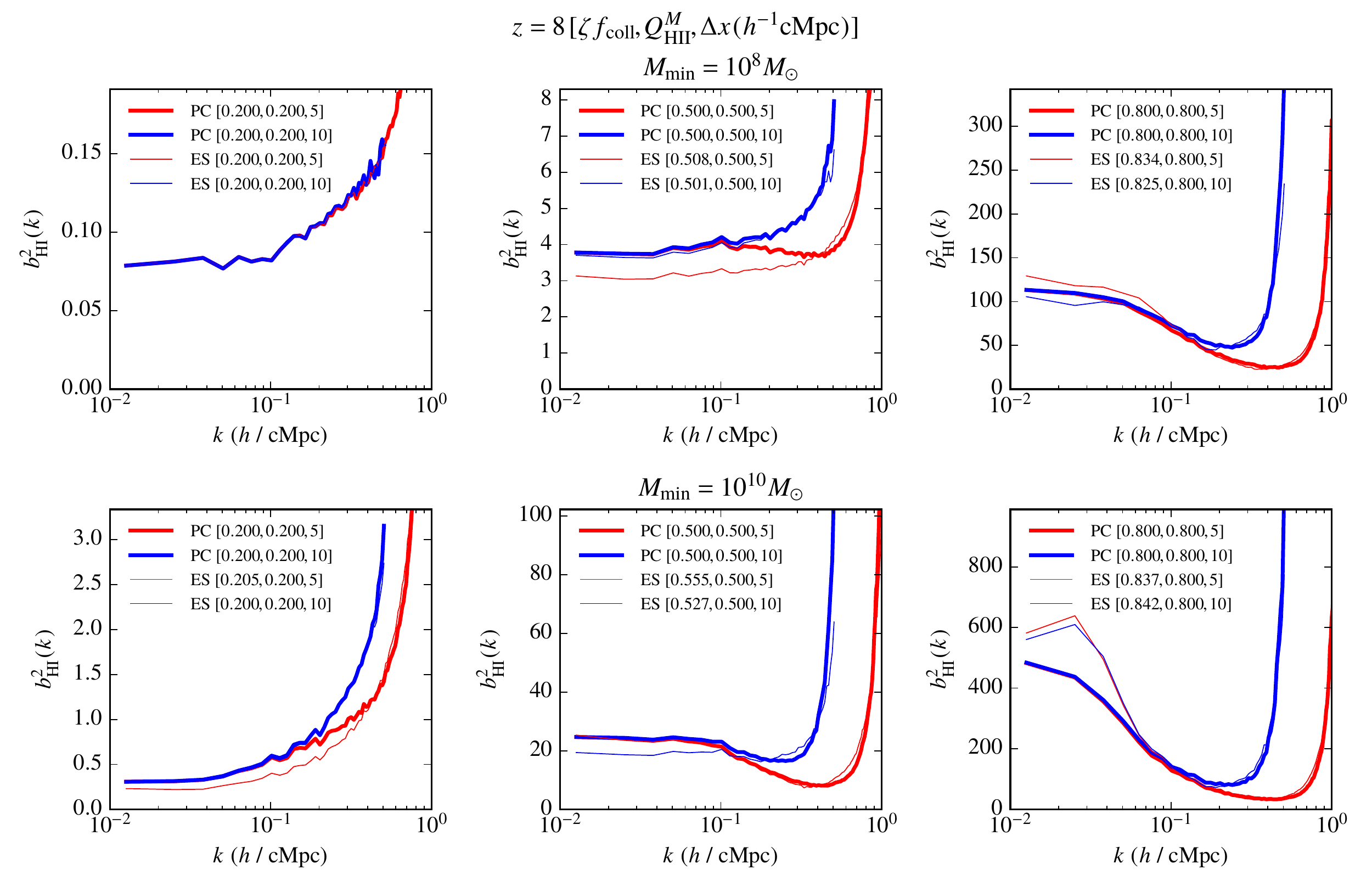}
\caption{
The bias $b_{\rm HI}^2(k) = P_{\rm HI}(k) / P(k)$ of the HI density fluctuations for the photon-conserving model. Different panels and curves are for different resolution and parameter values as indicated in the respective legends. For comparison, we show results obtained using the excursion set model by thin lines.
}
\label{fig:bias_6panels_onlyPC}
\end{figure*}

In \fig{fig:plotionizedmaps} we compare the simulated reionization maps produced by the two schemes. As before, we concentrate on $z = 8$, $M_{\rm min} = 10^8 \Msun$ and vary $\zeta$ to obtain different $Q_{\rm HII}^M$, as indicated above each panel. For this comparison, we use the sharp-$k$ filter in the ES models because it produces maps with less severe photon non-conservation and bias non-convergence. It is clear that the maps for the ES and PC models are identical when $Q_{\rm HII}^M = 0.2$ (left panels) as both the schemes conserve photons. In this case, the ionization field simply traces the underlying (smoothed) halo field. It is almost impossible to see any differences in the case $Q_{\rm HII}^M = 0.5$ (middle panels) where photon non-conservation has begun to set in for the ES model. One can see the differences more clearly in the right panel where $Q_{\rm HII}^M = 0.8$ and the photon non-conservation is $\sim 4\%$ in the ES case. We find that the ionized regions in the PC maps seem to be ``fragmented'' compared to the ES maps.

Finally, we consider the 21~cm power spectra produced by our PC model, again focusing on a specific redshift $z = 8$. The HI bias for the PC model for different model parameters and resolutions are shown in \fig{fig:bias_6panels_onlyPC}, the values of the relevant parameters can be read off from the respective legends. For comparison, we also show the corresponding results for the ES model by thin lines (these curves are identical to those shown in \fig{fig:bias_6panels_onlyES}). It is obvious that the large-scale bias is converged between the two resolutions for our PC model for all considered values of $Q_{\rm HII}^M$ and $M_{\rm min}$, as anticipated in \secn{subsec:photonNC->biasNC}. The bias for the ES model, in general, does not converge to the PC result for any of the resolutions considered in the Figure. In fact, the behaviour of the ES models, when compared to the converged PC results, is completely consistent with the arguments presented in \secn{subsec:photonNC->biasNC} and validated in Appendix~\ref{app:photonNC->biasNC}.\footnote{For example, for the smaller $M_{\rm min}$ maps, at intermediate and high ionized fractions (top middle and top right panels of \fig{fig:bias_6panels_onlyPC}) the coarser resolution ES results are closer to the PC results than the high-resolution ones. On the other hand, at higher $M_{\rm min}$ and at high ionization fraction (bottom right panel), the ES maps at \emph{both} resolutions give identical results that are different from the PC value, consistent with the argument in \secn{subsec:photonNC->biasNC} that the ES bubbles are always well resolved in this situation and are producing the wrong answer.} 

In particular, as we argued there, the ES method by construction produces the correct HI bias at sufficiently early stages of reionization (or at sufficiently coarse resolution) when the fraction of partially ionized cells is high. 
This is the case for the $\Delta x = 10 h^{-1}$cMpc results at smaller values of $\zeta f_{\rm coll}$ (top left, top middle and bottom left panels, compare bottom panel of \fig{fig:zetafcoll_by_Q_fixed_z}). The fact that our PC method matches each of these nearly perfectly, means that the PC answer has not only converged, but has converged to the correct value. In fact, we would argue that, being perfectly photon conserving and converged across resolution, it is the PC result itself which serves as the correct benchmark against which to compare the results of the ES method. 

\begin{figure*}
\includegraphics[width=0.6\textwidth]{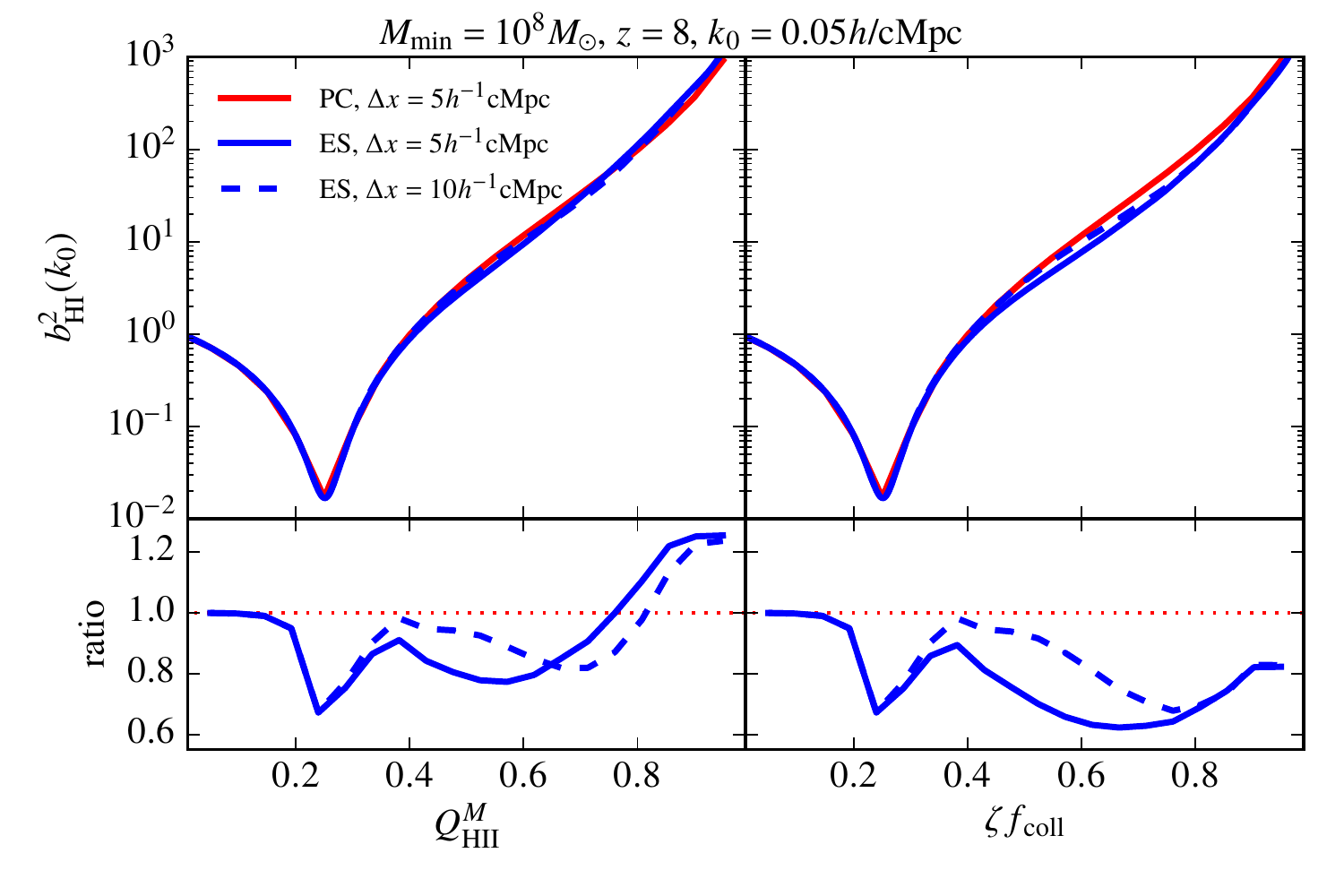}
\caption{
\emph{Top panels:} $b_{\rm HI}^2(k_0)$ for a representative $k = k_0 = 0.05~h/$cMpc as a function of $Q_{\rm HII}^M$ (left panel) and $\zeta f_{\rm coll}$ (right panel) for the photon-conserving model (PC, red lines) and the excursion set model with sharp-$k$ filter (ES, blue lines). The different values of the parameters are indicated in the legend. Note that we show the results for the PC model for only one resolution as the bias in this model is essentially perfectly converged at the scale of interest. \emph{Bottom panels:} The ratio of $b_{\rm HI}^2(k_0)$ for the ES model to that for the PC model. Horizontal dotted line indicates the value unity.
}
\label{fig:Pk_fixed_k_fixed_z_ES_PC}
\end{figure*}

To see the difference between the bias predicted in the two schemes in more detail, we plot $b^2_{\rm HI}(k_0)$ (where $k_0 = 0.05 h/\mbox{cMpc})$ as a function of $Q^M_{\rm HII}$ (top left panel) and $\zeta f_{\rm coll}$ (top right panel) in \fig{fig:Pk_fixed_k_fixed_z_ES_PC}, which show that the behaviour of $b^2_{\rm HI}(k_0)$ with varying $Q^M_{\rm HII}$ and $\zeta f_{\rm coll}$ is qualitatively similar in all the models.  The bottom panels show the ratio of $b^2_{\rm HI}(k_0)$ for the ES predictions to that for the PC model, and we see that the bias for the ES model can differ from the PC one by as much as $\sim 20$-$25\%$ for either resolution, especially at the intermediate and late stages of reionization where bubbles are well-resolved, despite the fact that level of photon non-conservation in the ES model is $\lesssim3$-$4\%$ at these stages (see \fig{fig:zetafcoll_by_Q_fixed_z}).\footnote{The prominent difference at $Q_{\rm HII}^M \sim 0.25$ is probably due to numerical effects, this is the stage where $b_{\rm HI}^2 \propto (1 - Q_{\rm HII}^M b_{\rm halo})^2$ (see Appendix \ref{app:bias_ES} for details) approaches zero, and the ratio of two small numbers can be numerically unstable.} In other words, \emph{the response of bias non-convergence to photon non-conservation in the ES models is quite large.}

We end this section by commenting on the relative efficiencies of the two schemes. The ES-based simulations are numerically quite inexpensive. For the kind of resolutions we use in this paper, the generation of one ionization map takes under a second on a single core on a laptop (assuming the density and halo fields have already been generated). In contrast, the PC scheme can take up to 4 hours for generating the default resolution maps, particularly towards the end stages of reionization where the overlap of bubbles is significant. It is thus clear that this method, though computationally faster than full radiative transfer simulations, is still not suitable for parameter estimation, e.g., using MCMC methods \citep{2015MNRAS.449.4246G}. Nevertheless, it is efficient enough to obtain maps for tens of different parameter values in a few days.

\section{Results for a realistic reionization model}
\label{sec:realistic}
The comparisons above focused on a fixed redshift. To study effects as a function of redshift, we consider a realistic but simple model of reionization. We assume $M_{\rm min} = 10^8 \Msun$ (i.e., only atomically cooled haloes contribute to ionizing photons) and take $\zeta$ to be constant. We assume that the globally averaged ionized mass fraction is given by $\mbox{min}[1, \zeta f_{\rm coll}(z)]$ with $f_{\rm coll}(z)$ being the one predicted by Sheth-Tormen mass function with the parameters adjusted as described previously. We can then calculate the Thomson scattering optical depth of the CMB photons as
\be
\tau_{\rm el} = \sigma_T c \int_0^{z_{\rm LSS}} \de t~\mbox{min}[1, \zeta f_{\rm coll}(z)]~n_b(z).
\ee
The value of $\zeta$ can be fixed by requiring the ionization history to produce a value of $\tau_{\rm el}$ consistent with latest observations \citep{2016A&A...596A.108P}. We use $\zeta = 7.5$, which produces $\tau_{\rm el} = 0.058$ with reionization completing at $z = 5.6$. Since we obtain the ionization maps from outputs of an $N$-body simulation, we show results only for the values of redshifts where the snapshot particle positions were stored. For epochs of relevance, we find that for $z = 10, 9, 8, 7, 6$, the analytically calculated ionized fractions are $\zeta f_{\rm coll}(z) = 0.160,   0.255,   0.394,  0.592,  0.862$ respectively. Note that, because of photon non-conservation, the ES-based ionization maps generated using this value of $\zeta$ will produce smaller $Q_{\rm HII}^M$ than what is predicted by the analytical formula.

\begin{figure}
\includegraphics[width=0.5\textwidth]{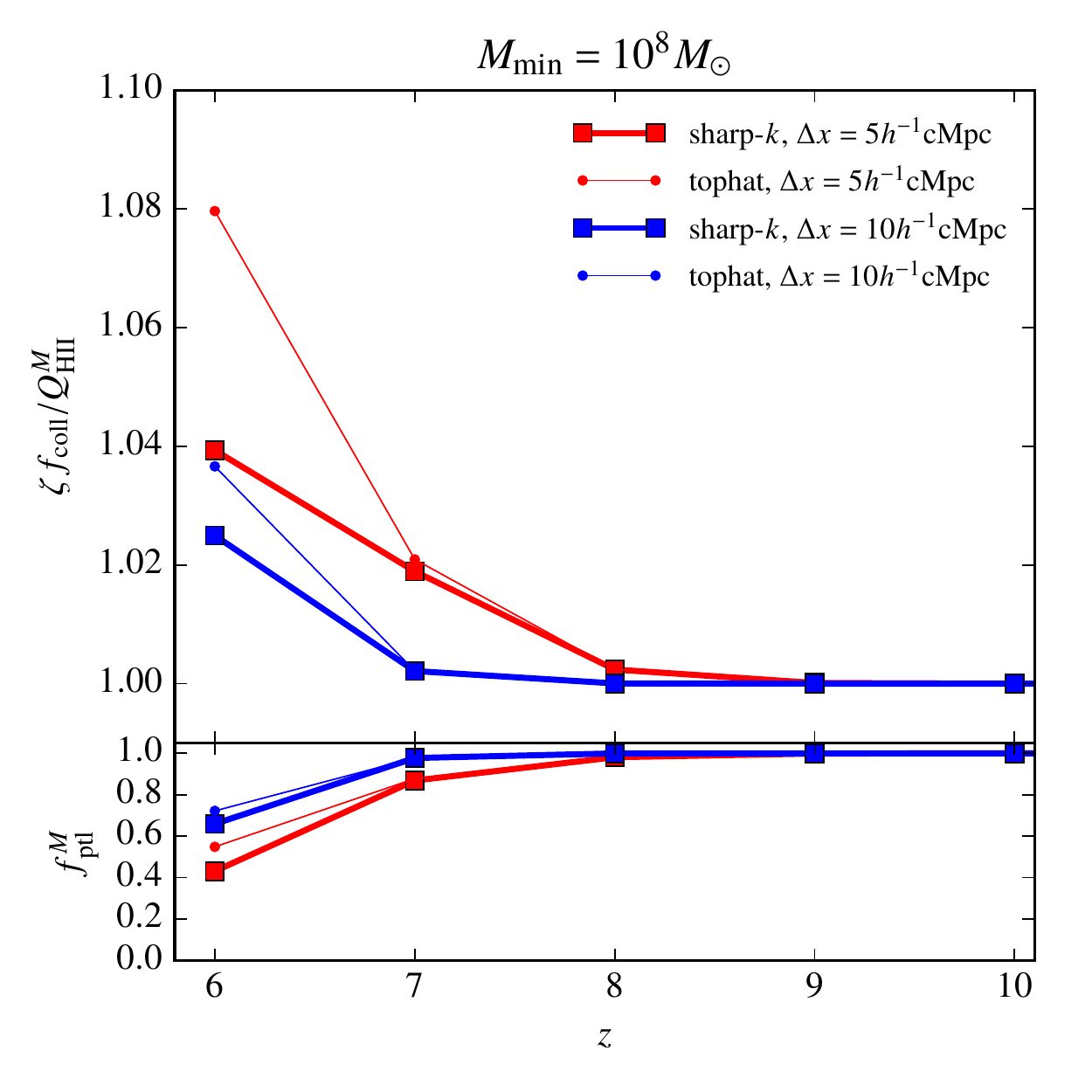}
\caption{
\emph{Top panel:} The ratio $\zeta f_{\rm coll} / Q^M_{\rm HII}$ as a function of $z$ for the fiducial reionization history for a neutral hydrogen field obtained using a conventional excursion set formalism. The thick lines are obtained using the sharp-$k$ filter while the thin lines are for the spherical tophat filter. The parameter values are indicated in the legend. \emph{Bottom panel:} The fraction $f_{\rm ptl}^M$ of ionized mass in partially ionized cells for the same parameter values. 
}
\label{fig:zetafcoll_by_Q_vary_z}
\end{figure}

The amount of photon non-conservation in the ES models as a function of $z$ for this reionization history is shown in the top panel of \fig{fig:zetafcoll_by_Q_vary_z} which is similar to the earlier \fig{fig:zetafcoll_by_Q_fixed_z}. It is clear that for the reionization history we have considered, there is almost no photon non-conservation at $z \gtrsim 8$ for the resolutions considered. This is because almost none of the bubbles are resolved and the ionized mass is mostly contained in partially ionized cells as can be seen from the bottom panel of the Figure.  The photon non-conservation is less than $4\%$ for the sharp-$k$ filter for the default resolution maps and can be less than $2\%$ for coarser resolutions. The corresponding non-conservation rises to about $8\%$ for the spherical tophat filter. The PC model, by construction, conserves photons at all redshifts, and hence we do not show it in this Figure.

\begin{figure}
\includegraphics[width=0.5\textwidth]{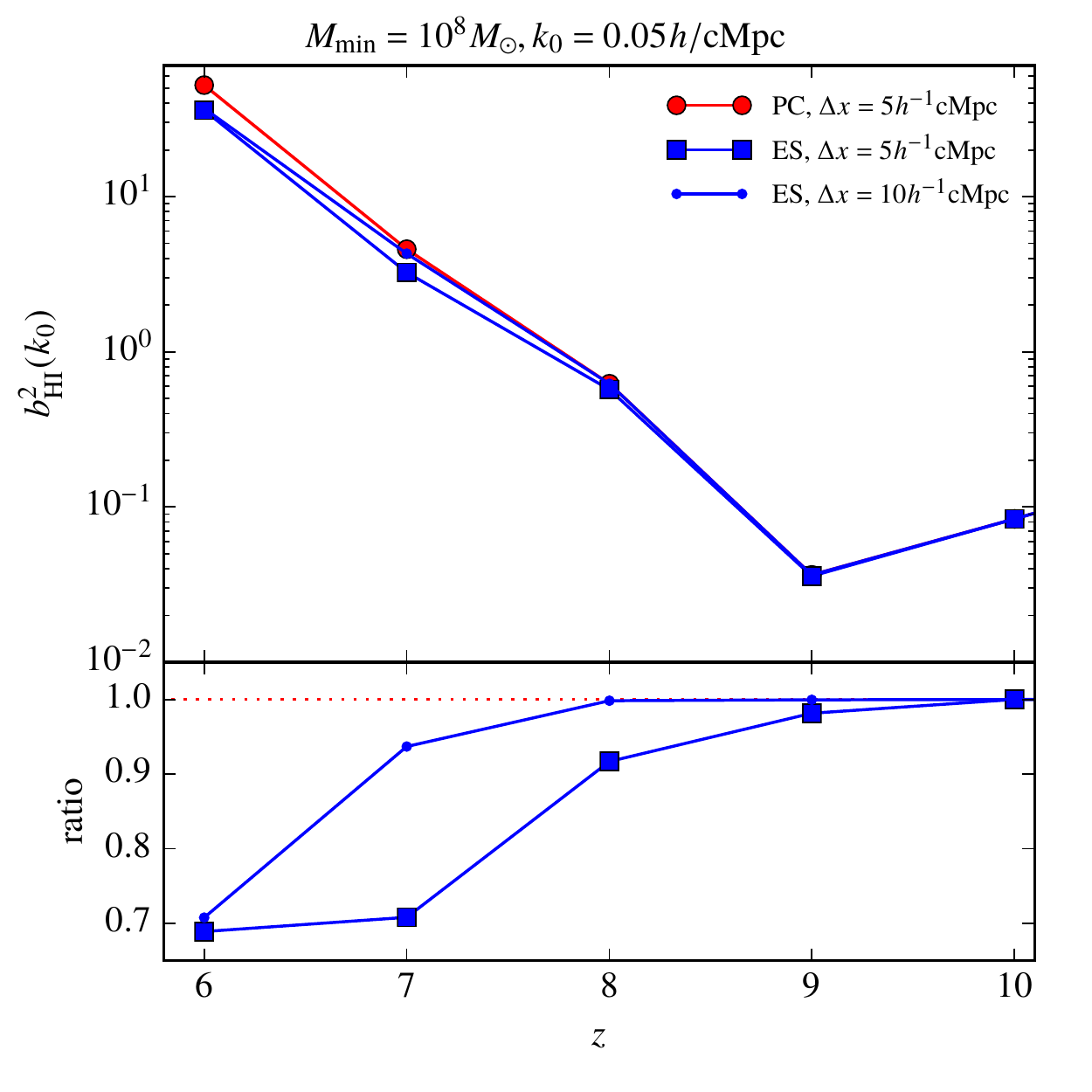}
\caption{
\emph{Top panel:} $b_{\rm HI}^2(k)$ for a representative $k = 0.05~h/$cMpc as a function of $z$ for the photon-conserving model (red lines) and the excursion set model with sharp-$k$ filter (blue lines) obtained for the fiducial reionization history. The resolutions of the maps used are indicated in the legend. Note that we show the results for the photon-conserving only for one resolution as the bias for the other resolution is the same at the scale of interest. \emph{Bottom panel:} The ratio of $b_{\rm HI}^2$ for the excursion set model to that for the photon-conserving model.
Horizontal line indicates the value unity.
}
\label{fig:Pk_fixed_k_vary_z_ES_PC}
\end{figure}

We next take a look at the behaviour of the bias $b_{\rm HI}^2(k_0)$ (where $k_0 = 0.05 h/\mbox{cMpc}$) as a function of $z$ for both the schemes in the top panel of \fig{fig:Pk_fixed_k_vary_z_ES_PC} (we only show results for the sharp-$k$ filter in the ES scheme). One should keep in mind that we have used the same value of $\zeta$ in all the models while computing the reionization history, hence the maps are normalized to same value of $\zeta f_{\rm coll}$. The qualitative behaviour of $b^2_{\rm HI}(k_0)$ for all the cases is similar. The difference in the bias between the two schemes can be seen from the bottom panel of the same Figure. The differences, as we see here, can be as large as $30\%$ particularly towards the end stages of reionization (whereas the corresponding photon non-conservation seen in \fig{fig:zetafcoll_by_Q_vary_z} at these redshifts is only $\sim2$-$4\%$). Thus the conclusions drawn for the $z = 8$ maps hold true for a realistic reionization history as well.

\section{Summary and Conclusions}
\label{sec:conclude}
In this paper, we revisited a previously known shortcoming of excursion set (ES) based semi-numerical models of reionization, namely, the fact that these models generically do not conserve photon number \citep{2007ApJ...654...12Z,2016MNRAS.460.1801P}. This problem \emph{per se} is relatively mild -- being at the $\sim$ few per cent level for state-of-the-art implementations -- and is probably not of great concern by itself, given the large astrophysical uncertainties involved in modelling the epoch of reionization. 

We have demonstrated, however, that this discrepancy of a few per cent in photon number conservation generically leads to a more severe problem, namely, that \emph{the large-scale 21~cm power spectrum predicted by ES models is a relatively strong function of the resolution chosen to generate the ionization maps}. E.g., the resolution-dependence of large-scale HI bias can be as large as $\sim25$-$30\%$ as compared to the fully converged value, even when the corresponding photon non-conservation is as small as $\sim4\%$ (compare Figures~\ref{fig:zetafcoll_by_Q_fixed_z} and~\ref{fig:Pk_fixed_k_fixed_z_ES_PC}, and also Figures~\ref{fig:Pk_fixed_k_vary_z_ES_PC} and~\ref{fig:zetafcoll_by_Q_vary_z}).

These differences may not be significant for interpreting the data from ongoing telescopes like the LOFAR, MWA and PAPER where the primary aim is to make a detection of the 21~cm signal and possibly put only some limits on parameter values \citep{2016MNRAS.455.4295G}. However, with the next generation of telescopes like the SKA (Phase 1) and HERA, one expects the errors on the power spectra to be $\lesssim 10\%$ at $k \lesssim 0.1/$cMpc \citep[see, e.g.,][]{2015aska.confE...1K}. Theoretical uncertainties of $\gtrsim 25\%$ could therefore significantly affect the interpretation of the data, making it important to understand the origin of this problem.

We traced this resolution-dependence of the large-scale HI bias to the fact that photon non-conservation in ES-based semi-numerical models is itself resolution-dependent, since it arises from a mixture of resolved bubbles which do not conserve photons in the ES method \citep[see][]{2016MNRAS.460.1801P} and partially ionized grid cells which are perfectly photon-conserving (see \secn{subsec:ES-photonNC}). This resolution-dependence of photon non-conservation leads to a resolution-dependence of the large-scale 21~cm power spectrum for maps generated both, at fixed ionized mass fraction $Q_{\rm HII}^M$ as well as at fixed ionizing efficiency $\zeta$ (\secn{subsec:photonNC->biasNC} and Appendix~\ref{app:photonNC->biasNC}).  Photon conservation and the convergence of the large-scale power spectrum can be ensured in the ES method only when the maps have so coarse a resolution that almost no ionized bubbles are resolved. 

\begin{figure}
\includegraphics[width=0.5\textwidth]{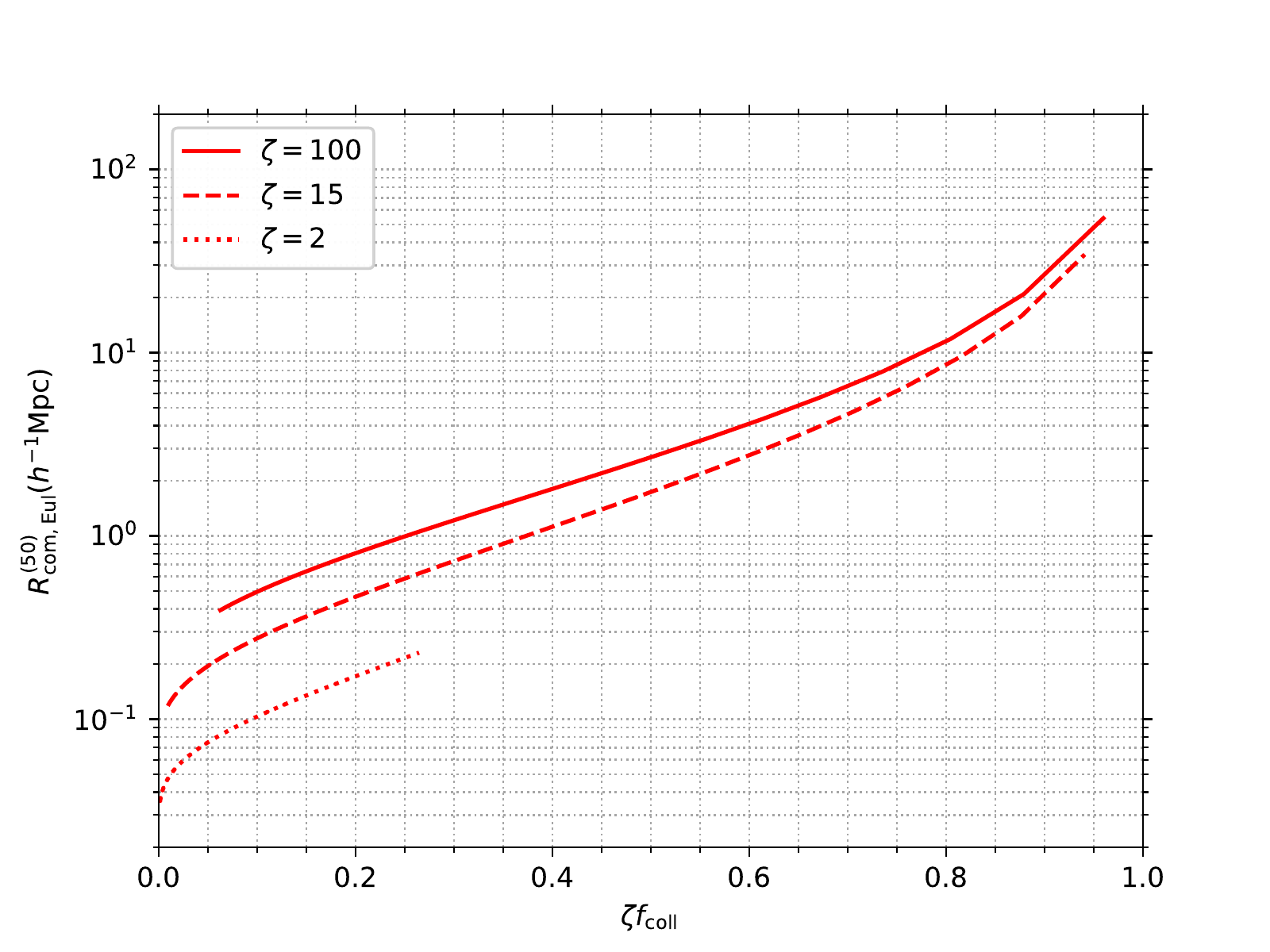}
\caption{
The median comoving Eulerian bubble radius $R_{\rm com,Eul}^{50}$, defined such that $50\%$ of the ionized mass is contained in bubbles of radius $<R_{\rm com,Eul}^{50}$. We show results computed using the model of \citet{2004ApJ...613....1F} with the Planck cosmology and three different values of $\zeta$ (the central value leads to a realistic $\tau_{\rm el}$), as a function of $\zeta f_{\rm coll}$. If one decides to produce ionization maps using an ES-based approach at some resolution $\Delta x$, then photon-conservation and bias-convergence are guaranteed only so long as $R_{\rm com,Eul}^{50} \ll \Delta x$. Thus, setting $R_{\rm com,Eul}^{50} = \Delta x$ in this plot gives an indication of the largest ionized fraction until which the ES-based approach will produce photon-conserving, bias-converged results.
}
\label{fig:cbs-zetafc}
\end{figure}

One might argue that a reasonable compromise is to therefore work with only low-resolution simulations, particularly if one is interested in only large-scale behaviour of the fluctuations. However, ionized bubbles grow as reionization progresses, so this resolution must be adjusted for the stage of reionization, i.e., it must be coarser for higher $Q_{\rm HII}^M$. E.g., consider the scale $R^{50}_{\rm Eul}$ defined such that 50\% of the ionized mass resides in bubbles of radii $< R^{50}_{\rm Eul}$. \fig{fig:cbs-zetafc} shows this quantity estimated using the analytical ES-based model of \citet[][FZH04]{2004ApJ...613....1F} as a function of $\zeta f_{\rm coll}$ for three different values of $\zeta$.\footnote{We show results for two extreme values of $\zeta$ surrounding a central value $\zeta=15$; the latter leads to a $\tau_{\rm el}$ consistent with the recent observational constraints. One should keep in mind that these analytical results may not be completely consistent with our semi-numerical approach, so these numbers should be treated as indicative only.} If we assume that one needs a grid size larger than $R^{50}_{\rm Eul}$ to be sufficiently photon-conserving and bias-converged, then we can conclude from the Figure that a resolution of $\Delta x = 5 h^{-1}$cMpc would allow us to follow the model only up to $Q_{\rm HII}^M \lesssim 0.7$, whereas using $\Delta x = 10 h^{-1}$cMpc would allow to reach slightly later $Q_{\rm HII}^M \lesssim 0.8$.

Restricting to low-resolution simulations would also imply that the bubble size distribution cannot be probed below a certain scale.\footnote{In this work we are restricted to somewhat low resolution simulations for a different reason. As discussed in \secn{sec:density_halo_fields} and Appendix \ref{app:halofield}, our prescription for assigning a collapsed mass fraction to the grid cells becomes inaccurate for too fine a grid size.} It is believed that this size distribution, in principle, contains information about the clustering of the ionizing sources, the underlying matter power spectrum, the stochasticity in the source population, etc. \citep{2006MNRAS.365..115F}. Topological features of the ionization field could also contain information regarding cosmological initial conditions \citep{2017MNRAS.466.2302B,2017arXiv171209195K,2018MNRAS.477.1984B}. All these interesting connections would remain obscured in the low-resolution maps thus making them somewhat ineffective while interpreting the 21~cm data.

With the point of view that the topology of ionized regions during reionization is interesting enough to be modelled accurately, we therefore proposed an explicitly photon-conserving (PC) approach to generate the ionization field, as an alternative to the standard ES method (\secn{sec:PC}). Our model is based on first generating bubbles around individual grid cells (or sources, as the case may be) and then dealing with the regions in the overlapped bubbles separately. Our method of distributing the photons in grid cells ensures photon conservation by construction, and produces a large-scale bias that is independent of the map resolution. As such, our PC model is the first example of a semi-numerical method that can serve as a valid benchmark for any ES-based technique as far as resolution-dependent effects are concerned.

At present, the main difficulty in replacing the ES model with our PC model is that the latter is computationally slower and hence is not suitable for parameter estimation. While ES-based models thus remain invaluable because of the flexibility they provide, one needs to be careful while interpreting the large-scale power spectrum particularly for higher resolution maps. In the future it will be useful to work on photon-conserving algorithms that are computationally faster and more efficient; this is the subject of work in progress. 

\section*{Acknowledgments}
The research of AP is supported by the Associateship Scheme of ICTP, Trieste and the Ramanujan Fellowship awarded by the Department of Science and Technology, Government of India. TRC acknowledges support from the Associateship Scheme of ICTP, Trieste and useful discussions with Raghunath Ghara. We thank the anonymous referee for a careful review of the paper.

\bibliographystyle{mnras}
\bibliography{ms,ms_AP}

\begin{thebibliography}{}
\makeatletter
\relax
\def\mn@urlcharsother{\let\do\@makeother \do\$\do\&\do\#\do\^\do\_\do\%\do\~}
\def\mn@doi{\begingroup\mn@urlcharsother \@ifnextchar [ {\mn@doi@}
  {\mn@doi@[]}}
\def\mn@doi@[#1]#2{\def\@tempa{#1}\ifx\@tempa\@empty \href
  {http://dx.doi.org/#2} {doi:#2}\else \href {http://dx.doi.org/#2} {#1}\fi
  \endgroup}
\def\mn@eprint#1#2{\mn@eprint@#1:#2::\@nil}
\def\mn@eprint@arXiv#1{\href {http://arxiv.org/abs/#1} {{\tt arXiv:#1}}}
\def\mn@eprint@dblp#1{\href {http://dblp.uni-trier.de/rec/bibtex/#1.xml}
  {dblp:#1}}
\def\mn@eprint@#1:#2:#3:#4\@nil{\def\@tempa {#1}\def\@tempb {#2}\def\@tempc
  {#3}\ifx \@tempc \@empty \let \@tempc \@tempb \let \@tempb \@tempa \fi \ifx
  \@tempb \@empty \def\@tempb {arXiv}\fi \@ifundefined
  {mn@eprint@\@tempb}{\@tempb:\@tempc}{\expandafter \expandafter \csname
  mn@eprint@\@tempb\endcsname \expandafter{\@tempc}}}

\bibitem[\protect\citeauthoryear{{Angulo}, {Baugh}, {Frenk}  \&
  {Lacey}}{{Angulo} et~al.}{2014}]{abfl14}
{Angulo} R.~E.,  {Baugh} C.~M.,  {Frenk} C.~S.,   {Lacey} C.~G.,  2014, \mn@doi
  [\mnras] {10.1093/mnras/stu1084}, \href
  {http://adsabs.harvard.edu/abs/2014MNRAS.442.3256A} {442, 3256}

\bibitem[\protect\citeauthoryear{{Bag}, {Mondal}, {Sarkar}, {Bharadwaj}  \&
  {Sahni}}{{Bag} et~al.}{2018}]{2018MNRAS.477.1984B}
{Bag} S.,  {Mondal} R.,  {Sarkar} P.,  {Bharadwaj} S.,   {Sahni} V.,  2018,
  \mn@doi [\mnras] {10.1093/mnras/sty714}, \href
  {http://adsabs.harvard.edu/abs/2018MNRAS.477.1984B} {477, 1984}

\bibitem[\protect\citeauthoryear{{Balaguera-Antol{\'{\i}}nez}, {Kitaura},
  {Pellerejo-Iba{\~n}ez}, {Zhao}  \& {Abel}}{{Balaguera-Antol{\'{\i}}nez}
  et~al.}{2018}]{balaguera-antolinez+18}
{Balaguera-Antol{\'{\i}}nez} A.,  {Kitaura} F.-S.,  {Pellerejo-Iba{\~n}ez} M.,
  {Zhao} C.,   {Abel} T.,  2018, preprint, \href
  {http://adsabs.harvard.edu/abs/2018arXiv180605870B} {} (\mn@eprint {arXiv}
  {1806.05870})

\bibitem[\protect\citeauthoryear{{Bandyopadhyay}, {Choudhury}  \&
  {Seshadri}}{{Bandyopadhyay} et~al.}{2017}]{2017MNRAS.466.2302B}
{Bandyopadhyay} B.,  {Choudhury} T.~R.,   {Seshadri} T.~R.,  2017, \mn@doi
  [\mnras] {10.1093/mnras/stw3347}, \href
  {http://adsabs.harvard.edu/abs/2017MNRAS.466.2302B} {466, 2302}

\bibitem[\protect\citeauthoryear{{Barkana} \& {Loeb}}{{Barkana} \&
  {Loeb}}{2001}]{2001PhR...349..125B}
{Barkana} R.,  {Loeb} A.,  2001, \mn@doi [\physrep]
  {10.1016/S0370-1573(01)00019-9}, \href
  {http://adsabs.harvard.edu/abs/2001PhR...349..125B} {349, 125}

\bibitem[\protect\citeauthoryear{{Barkana} \& {Loeb}}{{Barkana} \&
  {Loeb}}{2004}]{2004ApJ...609..474B}
{Barkana} R.,  {Loeb} A.,  2004, \mn@doi [\apj] {10.1086/421079}, \href
  {http://adsabs.harvard.edu/abs/2004ApJ...609..474B} {609, 474}

\bibitem[\protect\citeauthoryear{{Bond} \& {Myers}}{{Bond} \&
  {Myers}}{1996}]{bm96}
{Bond} J.~R.,  {Myers} S.~T.,  1996, \mn@doi [\apjs] {10.1086/192267}, \href
  {http://adsabs.harvard.edu/abs/1996ApJS..103....1B} {103, 1}

\bibitem[\protect\citeauthoryear{{Bond}, {Cole}, {Efstathiou}  \&
  {Kaiser}}{{Bond} et~al.}{1991}]{1991ApJ...379..440B}
{Bond} J.~R.,  {Cole} S.,  {Efstathiou} G.,   {Kaiser} N.,  1991, \mn@doi
  [\apj] {10.1086/170520}, \href
  {http://adsabs.harvard.edu/abs/1991ApJ...379..440B} {379, 440}

\bibitem[\protect\citeauthoryear{{Choudhury}}{{Choudhury}}{2009}]{2009CSci...97..841C}
{Choudhury} T.~R.,  2009, Current Science, \href
  {http://adsabs.harvard.edu/abs/2009CSci...97..841C} {97, 841}

\bibitem[\protect\citeauthoryear{{Choudhury}, {Haehnelt}  \&
  {Regan}}{{Choudhury} et~al.}{2009}]{2009MNRAS.394..960C}
{Choudhury} T.~R.,  {Haehnelt} M.~G.,   {Regan} J.,  2009, \mn@doi [\mnras]
  {10.1111/j.1365-2966.2008.14383.x}, \href
  {http://adsabs.harvard.edu/abs/2009MNRAS.394..960C} {394, 960}

\bibitem[\protect\citeauthoryear{{Furlanetto}, {Zaldarriaga}  \&
  {Hernquist}}{{Furlanetto} et~al.}{2004}]{2004ApJ...613....1F}
{Furlanetto} S.~R.,  {Zaldarriaga} M.,   {Hernquist} L.,  2004, \mn@doi [\apj]
  {10.1086/423025}, \href {http://adsabs.harvard.edu/abs/2004ApJ...613....1F}
  {613, 1}

\bibitem[\protect\citeauthoryear{{Furlanetto}, {McQuinn}  \&
  {Hernquist}}{{Furlanetto} et~al.}{2006a}]{2006MNRAS.365..115F}
{Furlanetto} S.~R.,  {McQuinn} M.,   {Hernquist} L.,  2006a, \mn@doi [\mnras]
  {10.1111/j.1365-2966.2005.09687.x}, \href
  {http://adsabs.harvard.edu/abs/2006MNRAS.365..115F} {365, 115}

\bibitem[\protect\citeauthoryear{{Furlanetto}, {Oh}  \& {Briggs}}{{Furlanetto}
  et~al.}{2006b}]{2006PhR...433..181F}
{Furlanetto} S.~R.,  {Oh} S.~P.,   {Briggs} F.~H.,  2006b, \mn@doi [\physrep]
  {10.1016/j.physrep.2006.08.002}, \href
  {http://adsabs.harvard.edu/abs/2006PhR...433..181F} {433, 181}

\bibitem[\protect\citeauthoryear{{Geil} \& {Wyithe}}{{Geil} \&
  {Wyithe}}{2008}]{2008MNRAS.386.1683G}
{Geil} P.~M.,  {Wyithe} J.~S.~B.,  2008, \mn@doi [\mnras]
  {10.1111/j.1365-2966.2008.13159.x}, \href
  {http://adsabs.harvard.edu/abs/2008MNRAS.386.1683G} {386, 1683}

\bibitem[\protect\citeauthoryear{{Ghara}, {Choudhury}  \& {Datta}}{{Ghara}
  et~al.}{2015a}]{2015MNRAS.447.1806G}
{Ghara} R.,  {Choudhury} T.~R.,   {Datta} K.~K.,  2015a, \mn@doi [\mnras]
  {10.1093/mnras/stu2512}, \href
  {http://adsabs.harvard.edu/abs/2015MNRAS.447.1806G} {447, 1806}

\bibitem[\protect\citeauthoryear{{Ghara}, {Datta}  \& {Choudhury}}{{Ghara}
  et~al.}{2015b}]{2015MNRAS.453.3143G}
{Ghara} R.,  {Datta} K.~K.,   {Choudhury} T.~R.,  2015b, \mn@doi [\mnras]
  {10.1093/mnras/stv1855}, \href
  {http://adsabs.harvard.edu/abs/2015MNRAS.453.3143G} {453, 3143}

\bibitem[\protect\citeauthoryear{{Ghara}, {Mellema}, {Giri}, {Choudhury},
  {Datta}  \& {Majumdar}}{{Ghara} et~al.}{2018}]{2018MNRAS.476.1741G}
{Ghara} R.,  {Mellema} G.,  {Giri} S.~K.,  {Choudhury} T.~R.,  {Datta} K.~K.,
  {Majumdar} S.,  2018, \mn@doi [\mnras] {10.1093/mnras/sty314}, \href
  {http://adsabs.harvard.edu/abs/2018MNRAS.476.1741G} {476, 1741}

\bibitem[\protect\citeauthoryear{{Gnedin}, {Kravtsov}  \& {Chen}}{{Gnedin}
  et~al.}{2008}]{2008ApJ...672..765G}
{Gnedin} N.~Y.,  {Kravtsov} A.~V.,   {Chen} H.-W.,  2008, \mn@doi [\apj]
  {10.1086/524007}, \href {http://adsabs.harvard.edu/abs/2008ApJ...672..765G}
  {672, 765}

\bibitem[\protect\citeauthoryear{{Greig} \& {Mesinger}}{{Greig} \&
  {Mesinger}}{2015}]{2015MNRAS.449.4246G}
{Greig} B.,  {Mesinger} A.,  2015, \mn@doi [\mnras] {10.1093/mnras/stv571},
  \href {http://adsabs.harvard.edu/abs/2015MNRAS.449.4246G} {449, 4246}

\bibitem[\protect\citeauthoryear{{Greig}, {Mesinger}  \& {Pober}}{{Greig}
  et~al.}{2016}]{2016MNRAS.455.4295G}
{Greig} B.,  {Mesinger} A.,   {Pober} J.~C.,  2016, \mn@doi [\mnras]
  {10.1093/mnras/stv2618}, \href
  {http://adsabs.harvard.edu/abs/2016MNRAS.455.4295G} {455, 4295}

\bibitem[\protect\citeauthoryear{{Hassan}, {Dav{\'e}}, {Finlator}  \&
  {Santos}}{{Hassan} et~al.}{2017}]{2017MNRAS.468..122H}
{Hassan} S.,  {Dav{\'e}} R.,  {Finlator} K.,   {Santos} M.~G.,  2017, \mn@doi
  [\mnras] {10.1093/mnras/stx420}, \href
  {http://adsabs.harvard.edu/abs/2017MNRAS.468..122H} {468, 122}

\bibitem[\protect\citeauthoryear{{Jenkins}, {Frenk}, {White}, {Colberg},
  {Cole}, {Evrard}, {Couchman}  \& {Yoshida}}{{Jenkins}
  et~al.}{2001}]{2001MNRAS.321..372J}
{Jenkins} A.,  {Frenk} C.~S.,  {White} S.~D.~M.,  {Colberg} J.~M.,  {Cole} S.,
  {Evrard} A.~E.,  {Couchman} H.~M.~P.,   {Yoshida} N.,  2001, \mn@doi [\mnras]
  {10.1046/j.1365-8711.2001.04029.x}, \href
  {http://adsabs.harvard.edu/abs/2001MNRAS.321..372J} {321, 372}

\bibitem[\protect\citeauthoryear{{Jiang} \& {van den Bosch}}{{Jiang} \& {van
  den Bosch}}{2014}]{jvdb14}
{Jiang} F.,  {van den Bosch} F.~C.,  2014, \mn@doi [\mnras]
  {10.1093/mnras/stu280}, \href
  {http://adsabs.harvard.edu/abs/2014MNRAS.440..193J} {440, 193}

\bibitem[\protect\citeauthoryear{{Kapahtia}, {Chingangbam}, {Appleby}  \&
  {Park}}{{Kapahtia} et~al.}{2017}]{2017arXiv171209195K}
{Kapahtia} A.,  {Chingangbam} P.,  {Appleby} S.,   {Park} C.,  2017, preprint,
  \href {http://adsabs.harvard.edu/abs/2017arXiv171209195K} {} (\mn@eprint
  {arXiv} {1712.09195})

\bibitem[\protect\citeauthoryear{{Koopmans} et~al.,}{{Koopmans}
  et~al.}{2015}]{2015aska.confE...1K}
{Koopmans} L.,  et~al., 2015, Advancing Astrophysics with the Square Kilometre
  Array (AASKA14), \href {http://adsabs.harvard.edu/abs/2015aska.confE...1K}
  {p.~1}

\bibitem[\protect\citeauthoryear{{Kulkarni}, {Choudhury}, {Puchwein}  \&
  {Haehnelt}}{{Kulkarni} et~al.}{2017}]{2017MNRAS.469.4283K}
{Kulkarni} G.,  {Choudhury} T.~R.,  {Puchwein} E.,   {Haehnelt} M.~G.,  2017,
  \mn@doi [\mnras] {10.1093/mnras/stx1167}, \href
  {http://adsabs.harvard.edu/abs/2017MNRAS.469.4283K} {469, 4283}

\bibitem[\protect\citeauthoryear{{Lin}, {Oh}, {Furlanetto}  \& {Sutter}}{{Lin}
  et~al.}{2016}]{2016MNRAS.461.3361L}
{Lin} Y.,  {Oh} S.~P.,  {Furlanetto} S.~R.,   {Sutter} P.~M.,  2016, \mn@doi
  [\mnras] {10.1093/mnras/stw1542}, \href
  {http://adsabs.harvard.edu/abs/2016MNRAS.461.3361L} {461, 3361}

\bibitem[\protect\citeauthoryear{{Majumdar}, {Mellema}, {Datta}, {Jensen},
  {Choudhury}, {Bharadwaj}  \& {Friedrich}}{{Majumdar}
  et~al.}{2014}]{2014MNRAS.443.2843M}
{Majumdar} S.,  {Mellema} G.,  {Datta} K.~K.,  {Jensen} H.,  {Choudhury} T.~R.,
   {Bharadwaj} S.,   {Friedrich} M.~M.,  2014, \mn@doi [\mnras]
  {10.1093/mnras/stu1342}, \href
  {http://adsabs.harvard.edu/abs/2014MNRAS.443.2843M} {443, 2843}

\bibitem[\protect\citeauthoryear{{McQuinn}, {Furlanetto}, {Hernquist}, {Zahn}
  \& {Zaldarriaga}}{{McQuinn} et~al.}{2005}]{2005ApJ...630..643M}
{McQuinn} M.,  {Furlanetto} S.~R.,  {Hernquist} L.,  {Zahn} O.,   {Zaldarriaga}
  M.,  2005, \mn@doi [\apj] {10.1086/432049}, \href
  {http://adsabs.harvard.edu/abs/2005ApJ...630..643M} {630, 643}

\bibitem[\protect\citeauthoryear{{Mesinger} \& {Furlanetto}}{{Mesinger} \&
  {Furlanetto}}{2007}]{2007ApJ...669..663M}
{Mesinger} A.,  {Furlanetto} S.,  2007, \mn@doi [\apj] {10.1086/521806}, \href
  {http://adsabs.harvard.edu/abs/2007ApJ...669..663M} {669, 663}

\bibitem[\protect\citeauthoryear{{Mesinger}, {Furlanetto}  \& {Cen}}{{Mesinger}
  et~al.}{2011}]{2011MNRAS.411..955M}
{Mesinger} A.,  {Furlanetto} S.,   {Cen} R.,  2011, \mn@doi [\mnras]
  {10.1111/j.1365-2966.2010.17731.x}, \href
  {http://adsabs.harvard.edu/abs/2011MNRAS.411..955M} {411, 955}

\bibitem[\protect\citeauthoryear{{Mo} \& {White}}{{Mo} \& {White}}{1996}]{mw96}
{Mo} H.~J.,  {White} S.~D.~M.,  1996, \mnras, \href
  {http://adsabs.harvard.edu/abs/1996MNRAS.282..347M} {282, 347}

\bibitem[\protect\citeauthoryear{{Monaco}, {Theuns}  \& {Taffoni}}{{Monaco}
  et~al.}{2002}]{pinocchio}
{Monaco} P.,  {Theuns} T.,   {Taffoni} G.,  2002, \mnras, 331, 587

\bibitem[\protect\citeauthoryear{{Paranjape} \& {Choudhury}}{{Paranjape} \&
  {Choudhury}}{2014}]{2014MNRAS.442.1470P}
{Paranjape} A.,  {Choudhury} T.~R.,  2014, \mn@doi [\mnras]
  {10.1093/mnras/stu911}, \href
  {http://adsabs.harvard.edu/abs/2014MNRAS.442.1470P} {442, 1470}

\bibitem[\protect\citeauthoryear{{Paranjape}, {Choudhury}  \&
  {Padmanabhan}}{{Paranjape} et~al.}{2016}]{2016MNRAS.460.1801P}
{Paranjape} A.,  {Choudhury} T.~R.,   {Padmanabhan} H.,  2016, \mn@doi [\mnras]
  {10.1093/mnras/stw1060}, \href
  {http://adsabs.harvard.edu/abs/2016MNRAS.460.1801P} {460, 1801}

\bibitem[\protect\citeauthoryear{{Patil} et~al.,}{{Patil}
  et~al.}{2017}]{2017ApJ...838...65P}
{Patil} A.~H.,  et~al., 2017, \mn@doi [\apj] {10.3847/1538-4357/aa63e7}, \href
  {http://adsabs.harvard.edu/abs/2017ApJ...838...65P} {838, 65}

\bibitem[\protect\citeauthoryear{{Planck Collaboration} et~al.,}{{Planck
  Collaboration} et~al.}{2014}]{2014A&A...571A..16P}
{Planck Collaboration} et~al., 2014, \mn@doi [\aap]
  {10.1051/0004-6361/201321591}, \href
  {http://adsabs.harvard.edu/abs/2014A%26A...571A..16P} {571, A16}

\bibitem[\protect\citeauthoryear{{Planck Collaboration} et~al.,}{{Planck
  Collaboration} et~al.}{2016}]{2016A&A...596A.108P}
{Planck Collaboration} et~al., 2016, \mn@doi [\aap]
  {10.1051/0004-6361/201628897}, \href
  {http://adsabs.harvard.edu/abs/2016A%26A...596A.108P} {596, A108}

\bibitem[\protect\citeauthoryear{{Santos}, {Ferramacho}, {Silva}, {Amblard}  \&
  {Cooray}}{{Santos} et~al.}{2010}]{2010MNRAS.406.2421S}
{Santos} M.~G.,  {Ferramacho} L.,  {Silva} M.~B.,  {Amblard} A.,   {Cooray} A.,
   2010, \mn@doi [\mnras] {10.1111/j.1365-2966.2010.16898.x}, \href
  {http://adsabs.harvard.edu/abs/2010MNRAS.406.2421S} {406, 2421}

\bibitem[\protect\citeauthoryear{{Scoccimarro} \& {Sheth}}{{Scoccimarro} \&
  {Sheth}}{2002}]{ss02}
{Scoccimarro} R.,  {Sheth} R.~K.,  2002, \mnras, 329, 629

\bibitem[\protect\citeauthoryear{{Seehars}, {Paranjape}, {Witzemann},
  {Refregier}, {Amara}  \& {Akeret}}{{Seehars}
  et~al.}{2016}]{2016JCAP...03..001S}
{Seehars} S.,  {Paranjape} A.,  {Witzemann} A.,  {Refregier} A.,  {Amara} A.,
  {Akeret} J.,  2016, \mn@doi [\jcap] {10.1088/1475-7516/2016/03/001}, \href
  {http://adsabs.harvard.edu/abs/2016JCAP...03..001S} {3, 001}

\bibitem[\protect\citeauthoryear{{Sheth} \& {Lemson}}{{Sheth} \&
  {Lemson}}{1999a}]{sl99a}
{Sheth} R.~K.,  {Lemson} G.,  1999a, \mn@doi [\mnras]
  {10.1046/j.1365-8711.1999.02378.x}, \href
  {http://adsabs.harvard.edu/abs/1999MNRAS.304..767S} {304, 767}

\bibitem[\protect\citeauthoryear{{Sheth} \& {Lemson}}{{Sheth} \&
  {Lemson}}{1999b}]{sl99b}
{Sheth} R.~K.,  {Lemson} G.,  1999b, \mn@doi [\mnras]
  {10.1046/j.1365-8711.1999.02477.x}, \href
  {http://adsabs.harvard.edu/abs/1999MNRAS.305..946S} {305, 946}

\bibitem[\protect\citeauthoryear{{Sheth} \& {Tormen}}{{Sheth} \&
  {Tormen}}{1999}]{1999MNRAS.308..119S}
{Sheth} R.~K.,  {Tormen} G.,  1999, \mn@doi [\mnras]
  {10.1046/j.1365-8711.1999.02692.x}, \href
  {http://adsabs.harvard.edu/abs/1999MNRAS.308..119S} {308, 119}

\bibitem[\protect\citeauthoryear{{Sheth} \& {Tormen}}{{Sheth} \&
  {Tormen}}{2002}]{2002MNRAS.329...61S}
{Sheth} R.~K.,  {Tormen} G.,  2002, \mn@doi [\mnras]
  {10.1046/j.1365-8711.2002.04950.x}, \href
  {http://adsabs.harvard.edu/abs/2002MNRAS.329...61S} {329, 61}

\bibitem[\protect\citeauthoryear{{Springel}}{{Springel}}{2005}]{2005MNRAS.364.1105S}
{Springel} V.,  2005, \mn@doi [\mnras] {10.1111/j.1365-2966.2005.09655.x},
  \href {http://adsabs.harvard.edu/abs/2005MNRAS.364.1105S} {364, 1105}

\bibitem[\protect\citeauthoryear{{Thomas} \& {Zaroubi}}{{Thomas} \&
  {Zaroubi}}{2011}]{2011MNRAS.410.1377T}
{Thomas} R.~M.,  {Zaroubi} S.,  2011, \mn@doi [\mnras]
  {10.1111/j.1365-2966.2010.17525.x}, \href
  {http://adsabs.harvard.edu/abs/2011MNRAS.410.1377T} {410, 1377}

\bibitem[\protect\citeauthoryear{{Thomas} et~al.,}{{Thomas}
  et~al.}{2009}]{2009MNRAS.393...32T}
{Thomas} R.~M.,  et~al., 2009, \mn@doi [\mnras]
  {10.1111/j.1365-2966.2008.14206.x}, \href
  {http://adsabs.harvard.edu/abs/2009MNRAS.393...32T} {393, 32}

\bibitem[\protect\citeauthoryear{{Zahn}, {Lidz}, {McQuinn}, {Dutta},
  {Hernquist}, {Zaldarriaga}  \& {Furlanetto}}{{Zahn}
  et~al.}{2007}]{2007ApJ...654...12Z}
{Zahn} O.,  {Lidz} A.,  {McQuinn} M.,  {Dutta} S.,  {Hernquist} L.,
  {Zaldarriaga} M.,   {Furlanetto} S.~R.,  2007, \mn@doi [\apj]
  {10.1086/509597}, \href {http://adsabs.harvard.edu/abs/2007ApJ...654...12Z}
  {654, 12}

\bibitem[\protect\citeauthoryear{{Zahn}, {Mesinger}, {McQuinn}, {Trac}, {Cen}
  \& {Hernquist}}{{Zahn} et~al.}{2011}]{2011MNRAS.414..727Z}
{Zahn} O.,  {Mesinger} A.,  {McQuinn} M.,  {Trac} H.,  {Cen} R.,   {Hernquist}
  L.~E.,  2011, \mn@doi [\mnras] {10.1111/j.1365-2966.2011.18439.x}, \href
  {http://adsabs.harvard.edu/abs/2011MNRAS.414..727Z} {414, 727}

\makeatother
\end{thebibliography}

\appendix

\section{Semi-analytical halo field}
\label{app:halofield}
The idea of generating halo locations and properties using only the coarse-grained properties of the dark matter field has a long history \citep{bm96,pinocchio,ss02} with renewed interest in recent times for the low-redshift mass function \citep[see, e.g.,][and references therein]{balaguera-antolinez+18}. 
One such algorithm usually used in high-redshift simulations is based on a hybrid prescription introduced by \citet{2004ApJ...609..474B}. This method involves computing the collapsed fraction $f_{\rm coll}$ in each grid cell using the conditional Press-Schechter mass function \citep{1991ApJ...379..440B}, and then employ a global scaling to match the mean $f_{\rm coll}$ to the value obtained from the mass function of \citet{1999MNRAS.308..119S} which has a better agreement with numerical simulations.

Ideally, one would use a calibration of the conditional mass function needed above directly from high-resolution $N$-body simulations \citep{abfl14}. In this work, we use the approximation discussed by \citet{2016JCAP...03..001S} which uses the conditional mass function obtained from the ellipsoidal collapse model \citep{2002MNRAS.329...61S}. Consider a region (say, a cell in a simulation volume) which has a mass $M_{\rm cell}$ and let $\delta_{{\rm L}, 0}$ be the corresponding density contrast in the initial conditions linearly extrapolated to $z = 0$. The conditional mass function in such a region is given by
\be
n (M | M_{\rm cell}, \delta_{\rm L, 0}) = \f{\bar{\rho}_m}{M}~ f(s | s_{\rm cell}, \delta_{\rm L, 0})~\left|\f{\de s}{\de M}\right|,
\ee
where $s \equiv \sigma^2(M)$ is the variance of the linearly extrapolated density scale at a mass scale $M$ and $s_{\rm cell} = \sigma^2(M_{\rm cell})$. The quantity $f(s | s_{\rm cell}, \delta_{\rm L, 0})$ can be calculated using the ES formalism and for the the ellipsoidal collapse barrier it is given by \citep{2002MNRAS.329...61S}
\be
f(s | s_{\rm cell}, \delta_{\rm L, 0}) = \f{1}{\sqrt{2 \pi}}~\f{\left|T(s | s_{\rm cell})\right|}{(s - s_{\rm cell})^{3/2}}~ \exp\left[-\f{[B(s) - \delta_{\rm L, 0}]^2}{2 (s - s_{\rm cell})}\right],
\label{eq:ellipsoidal_fs}
\ee
where
\be
B(s) = \sqrt{a}~\delta_c(z) \left[1 + \f{\beta}{(a \nu)^{\alpha}} \right]
\label{eq:barrier}
\ee
is the barrier corresponding to the ellipsoidal collapse and $T(s | s_{\rm cell})$ is the sum of the first few (six) terms of the Taylor series expansion of $B(s) -\delta_{\rm L, 0}$ around $s = s_{\rm cell}$
\be
T(s | s_{\rm cell}) = \sum_{n = 0}^5 \f{(s_{\rm cell}-s)^n}{n!}~\f{\del^n [B(s) - \delta_{\rm L, 0}]}{\del s^n}.
\ee

\begin{figure}
\includegraphics[width=0.5\textwidth]{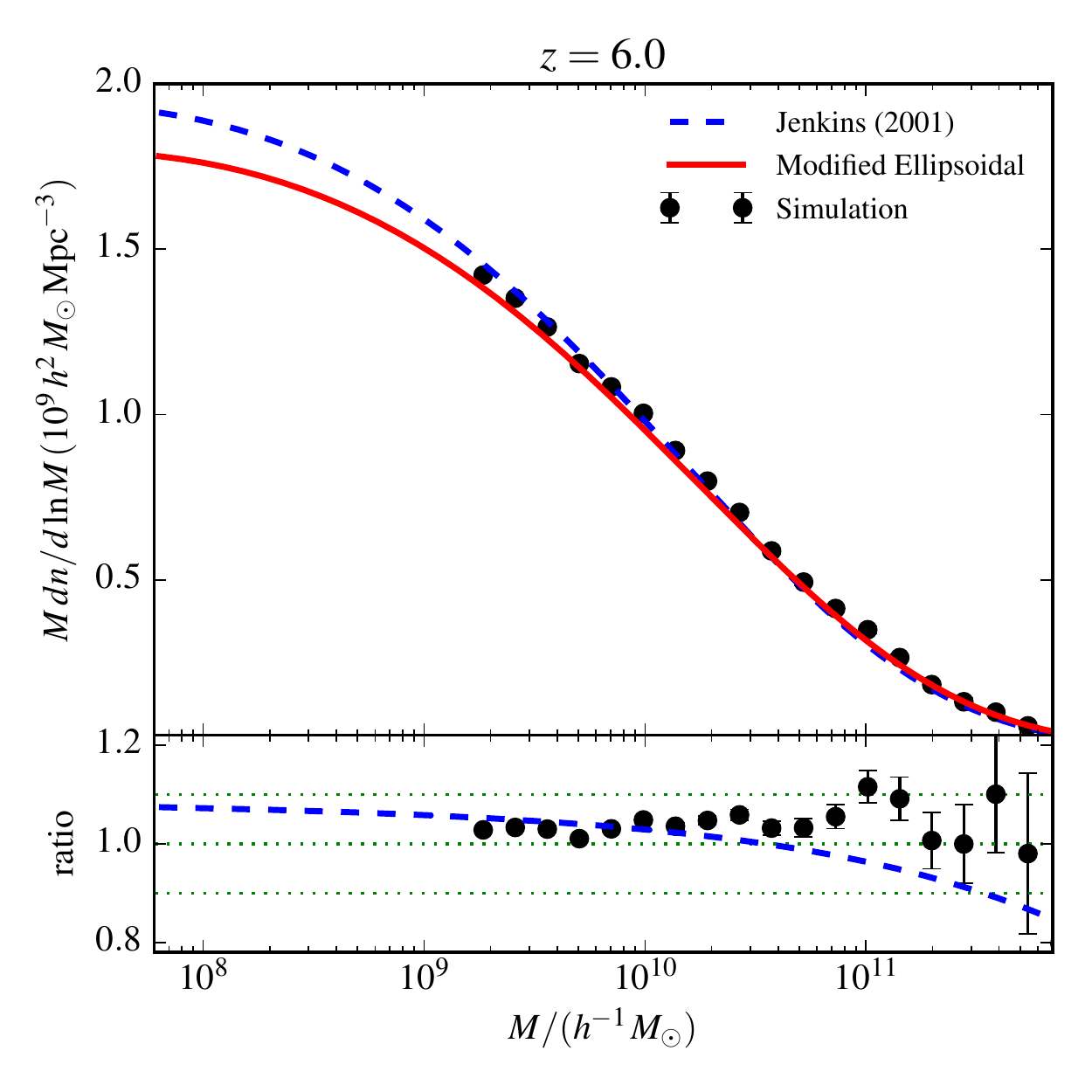}
\caption{
Comparison of the mass function parametrization used in this work (solid red) with the mass function obtained from an $N$-body simulation (symbols with error bars). Since the simulation cannot probe low-mass haloes (see text for the details), we also show the fitting function from \citet[][dashed blue]{2001MNRAS.321..372J}. The \emph{top panel} shows the unconditional mass fraction in units of $10^9\Mh/(\Mpch)^3$ and the \emph{bottom panel} shows the ratio of the various results with our parametrization. We see that all the results agree at the $\sim 10\%$ level for masses $>10^8\Msun$. Results are shown for $z=6$, with qualitatively similar comparisons obtained at other redshifts in the range $6\lesssim z\lesssim 15$. 
}
\label{fig:plot_halo_mass_func}
\end{figure}

In the expression for the barrier, i.e., \eqn{eq:barrier} above, $\delta_c(z) = 1.686 / D(z)$ is the critical density at $z$ and $\nu \equiv \delta^2_c(z) / s$, with $D(z)$ the linear theory growth factor normalized to unity at $z=0$. The values of the parameters appearing in the same expression, as suggested by \citet{2002MNRAS.329...61S}, are $\alpha = 0.615$ and $\beta = 0.485$ (as predicted by the ellipsoidal collapse dynamics) and $a = 0.75$ (obtained by fitting the simulations available at that time). We find, however, that the \emph{unconditional} mass function corresponding to the barrier in \eqn{eq:barrier} is a better fit to the mass function of Friends-of-Friends (FoF) haloes  obtained from $N$-body simulation at redshifts of our interest (i.e., $6 \lesssim z \lesssim 15$) if we change the values of the parameters to $\alpha = 0.7$, $\beta = 0.4$ and $a = 0.67$. 
To check this, we run a GADGET-2 $N$-body simulation with $1200^3$ particles in a volume $(100 h^{-1}~{\rm cMpc})^3$ which allows us to identify FoF haloes of masses $\gtrsim 2.3 \times 10^9 \Msun$ (assuming a minimum of 32 particles for a group to be labelled as a halo). 
As we see in \fig{fig:plot_halo_mass_func}, the adjusted parameters provide a $\sim10\%$ match with the mass function measured in our $N$-body simulation, as well as the fitting function of \citet{2001MNRAS.321..372J}, which is adequate for our purposes. 

Given the procedure for calculating the conditional mass function, the steps of generating the halo field required for our calculations are as follows:

\begin{itemize}

\item Given the particle positions in the simulation box, we use the cloud-in-cell (CIC) algorithm to generate a smooth density field in a uniform cubical grid. Let $\Delta x$ be the length of the grid cell. The mass contained in a grid cell located at a position $\mathbf{x}$ is then
\be
M_{\rm cell}(\mathbf{x}) = [1 + \delta_{\rm NL}(\mathbf{x})]~\bar{\rho}_m~(\Delta x)^3,
\ee
where $\delta_{\rm NL}(\mathbf{x})$ is the (non-linear) density contrast averaged over the cell.

\item To compute the quantity $\delta_{\rm L, 0}$ for the cell, we use the spherical collapse approximation to relate the non-linear density contrast to the linear one. These relations are given by the parametric equations \citep{mw96}
\be
\begin{array}{ll}
1 + \delta_{\rm NL}  =  \displaystyle\f{9}{2} \f{(\theta - \sin\theta)^2}{(1 - \cos \theta)^3}, &\delta_{\rm L}  =  \displaystyle\f{3 \times 6^{2/3}}{20} (\theta - \sin \theta)^{2/3},
\\
& \mbox{ for } \delta_{\rm NL} > 0,
\\
\\
1 + \delta_{\rm NL}  =  \displaystyle\f{9}{2} \f{(\sinh \theta - \theta)^2}{(\cosh \theta - 1)^3}, &\delta_{\rm L}  =  - \displaystyle\f{3 \times 6^{2/3}}{20} (\sinh \theta - \theta)^{2/3},
\\
& \mbox{ for } \delta_{\rm NL} < 0.
\\
\\
\end{array}
\label{eq:delta_NL_delta_L}
\ee
We numerically invert these relations for $\delta_{\rm L}$ in terms of $\delta_{\rm NL}$ to finally obtain $\delta_{\rm L,0} = \delta_{\rm L} / D(z)$.

\item Once we have the values of $\delta_{\rm L, 0}$ and $M_{\rm cell}$  for a cell, we can compute the conditional mass function $f(s | s_{\rm cell}, \delta_{{\rm L}, 0}$) using \eqn{eq:ellipsoidal_fs}. The collapse fraction above a mass $M_{\rm min}$ in the cell is given by
\be
f_{\rm coll}(\mathbf{x}) = \int_{s_{\rm cell}}^{s_{\rm min}} \de s~f(s | s_{\rm cell}, \delta_{{\rm L}, 0}),
\label{eq:fcoll}
\ee
where $s_{\rm min} = \sigma^2(M_{\rm min})$.

\end{itemize}

There a few further points to be discussed regarding our implementation of the above method. The first is the choice of the grid size $\Delta x$. Ideally one would like to calculate the collapse fraction in grid cells as small as that allowed by the original $N$-body simulation (set by the force softening scale). However, choosing too small a $\Delta x$ produces a $\delta_{\rm L}$ distribution, obtained using \eqn{eq:delta_NL_delta_L}, that is highly non-Gaussian, which is incorrect since the initial conditions of the simulation sampled a Gaussian random field. This happens mainly because the spherical approximation is not valid at very small scales where tidal effects become important, particularly in very high-density environments; i.e., the approximation works well when $\delta_{\rm NL}$ itself is not highly non-linear. A possible way of quantifying how well the approximation works is to calculate the skewness and excess kurtosis of the generated $\delta_{\rm L}$ distribution. For a representative redshift $z = 8$, we find that for our simulation box, the values of the skewness for $\Delta x = 2.5, 5.0, 10.0 h^{-1}$cMpc are $-0.25, -0.18, -0.12$, respectively. The excess kurtosis for the three cases is $\leq 0.08$. Clearly the approximation will be better if one uses a coarser resolution, however, one should keep in mind that it would not be possible to track the sizes of the ionized bubbles if the grid size is larger than the characteristic bubble radius. Since the characteristic bubble radius is $\lesssim 10 h^{-1}$cMpc when the average neutral fraction is $\sim 0.5$ \citep{2004ApJ...613....1F,2014MNRAS.442.1470P,2016MNRAS.461.3361L}, the default resolution we choose to work with is $\Delta x = 5.0 h^{-1}$cMpc.

Note that the collapse fraction as given by \eqn{eq:fcoll} represents the \emph{mean} value of the quantity. In the actual simulation, there would be scatter in the value of $f_{\rm coll}$ for the same $M_{\rm cell}$ and $\delta_{\rm NL}$. Although this scatter is non-trivial for halo masses approaching the cell size due to mass conservation, for large enough cell sizes it is possible to account for the scatter at a given redshift by simply Poisson-sampling the halo mass function \citep{sl99a,sl99b,2016JCAP...03..001S}. For multiple redshifts, one should in principle generate merger trees to avoid artificial appearance and disappearance of haloes as the density evolves \citep[see, e.g.][for a recent comparison of techniques]{jvdb14}. In this work, for simplicity, we completely ignore this scatter and assign only the mean value of $f_{\rm coll}$ to each grid cell. We have verified, using a Poisson-sampling version of our code, that the main conclusions of this paper do not depend on this choice.

\section{Behaviour of the large-scale bias in excursion set based models}
\label{app:bias_ES}

In this Appendix, we present some insights into the behaviour of the large-scale HI bias for the ES model and present some details on the connection between photon non-conservation and bias non-convergence.

\subsection{Behaviour of large-scale HI and HII bias}

\begin{figure}
\includegraphics[width=0.5\textwidth]{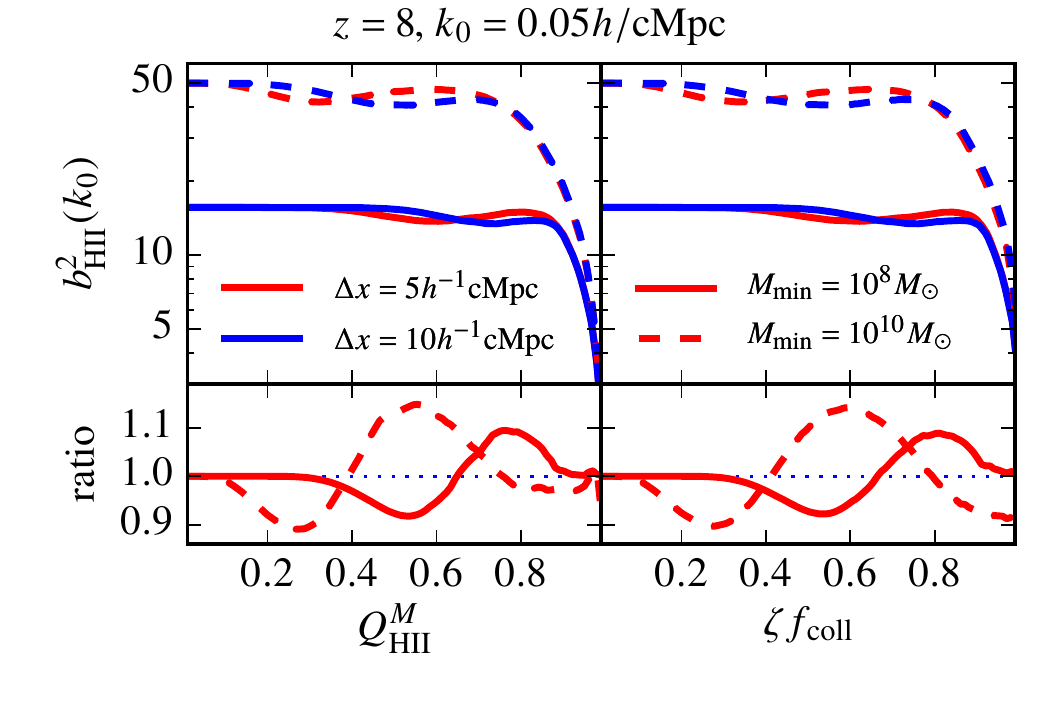}
\caption{
\emph{Top panels:} $b_{\rm HII}^2(k)$ for a representative $k = 0.05~h/$cMpc as a function of $Q_{\rm HII}^M$ (left panel) and $\zeta f_{\rm coll}$ (right panel) for the excursion set model with sharp-$k$ filter. The different values of the parameters are indicated in the legend. \emph{Bottom panels:} The ratio of $b_{\rm HII}^2$ for $\Delta x = 5 h^{-1}$cMpc to that for $\Delta x = 10 h^{-1}$cMpc for the same parameter values.
}
\label{fig:Pk_ionized_fixed_k_fixed_z}
\end{figure}

\begin{figure*}
\includegraphics[width=0.8\textwidth]{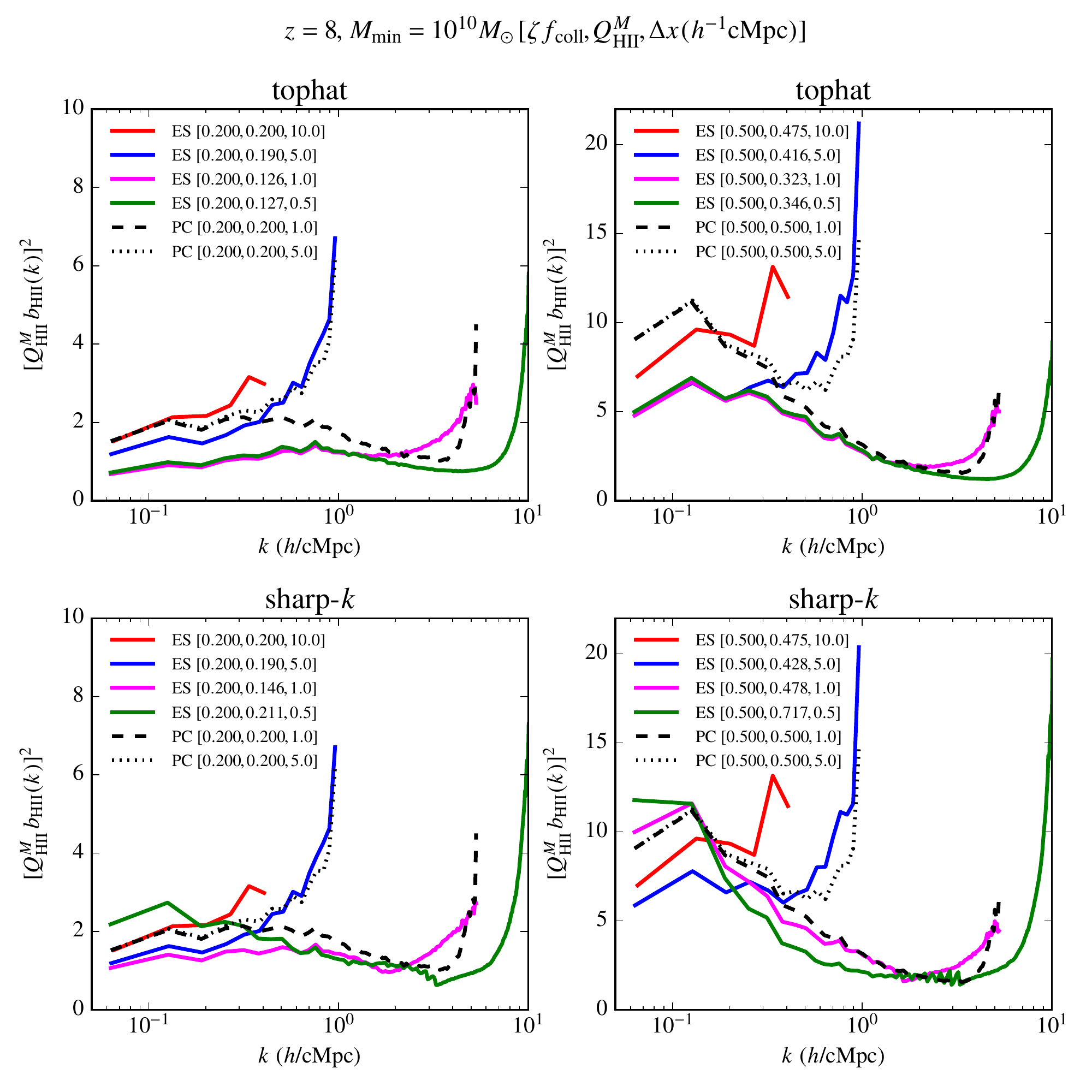}
\caption{
Convergence study of large-scale HII bias in the ES model at fixed ionizing efficiency $\zeta$. Solid curves in each panel show measurements of $(Q_{\rm HII}^Mb_{\rm HII})^2$ for maps generated using different resolutions but the same dark matter and halo field. (For this plot, we use haloes from our $N$-body simulation.) The \emph{top (bottom) row} shows results for the ES model with the spherical tophat (sharp-$k$) filter. Dashed and dotted curves (identical across top and bottom rows) show results for our PC model for two different resolutions. The \emph{left} and \emph{right panels} show results for two different values of $\zeta f_{\rm coll}$. The large-scale bias in the spherical tophat ES model converges at high resolution, but to the wrong answer, while the situation is more complex for the sharp-$k$ ES model. Note also the high level of photon non-convergence in the ES models at high resolution. See text for a discussion.
}
\label{fig:unnormalized_bias_ionized_4panels_haloes}
\end{figure*}

It is actually more convenient to understand the behaviour of $b_{\rm HI}(k)$ in terms of the \emph{ionized} overdensity field
\be
\Delta_{\rm HII}(\mathbf{x}) = \f{x_{\rm HII}(\mathbf{x}) [1 + \delta_{\rm NL}(\mathbf{x})]}{Q_{\rm HII}^M},
\ee
which is related to the HI overdensity by
\be 
\Delta_{\rm HI}(\mathbf{x}) = \f{1 + \delta_{\rm NL} - Q_{\rm HII}^M \Delta_{\rm HII}(\mathbf{x})}{1 - Q_{\rm HII}^M}.
\ee
At large scales, when the fluctuations are linear, we can assume that the bias is scale-free and deterministic. Under such assumptions, we can relate the HII bias to $b_{\rm HI}$ as
\be
b_{\rm HI} = \f{1 - Q_{\rm HII}^M b_{\rm HII}}{1 - Q_{\rm HII}^M}.
\label{eq:b_HI_b_HII}
\ee
Note that this assumption of a scale-free, deterministic bias will not hold even at the largest scales we are interested in towards the late stages of reionization when the bubble radii become comparable to these scales.

In \fig{fig:Pk_ionized_fixed_k_fixed_z}, we show $b_{\rm HII}^2(k_0)$ (where $k_0 = 0.05 h/{\rm Mpc}$) as a function of $Q_{\rm HII}^M$ (top left panel) and $\zeta f_{\rm coll}$ (top right panel) which are obtained using the ES-based semi-numerical simulations for different parameter values. The bottom panels of the Figure show the ratio of $b_{\rm HII}^2(k_0)$ for the default resolution ($\Delta x = 5 h^{-1}$cMpc) to that for the coarser one ($\Delta x = 10 h^{-1}$cMpc). 

In the early stages of reionization, we see that $b_{\rm HII}$ is independent of $Q_{\rm HII}^M$ and $\zeta f_{\rm coll}$. In fact, one can prove that the ES model prediction for $b_{\rm HII}$ (and hence for $b_{\rm HI}$) in this case is the correct answer by construction. This is because, at these stages, none of the bubbles are resolved and the ES model sets the ionized fraction of essentially each cell to be precisely equal to the correct answer in this limit, namely $\zeta f_{\rm coll}(\mathbf{x})$. The HII overdensity (assuming we are only interested in large scales) can be written as 
\begin{align}
Q_{\rm HII}^M \Delta_{\rm HII}(\mathbf{x}) &= x_{\rm HII}(\mathbf{x})\left(1+\delta_{\rm NL}(\mathbf{x})\right) \notag\\
&\approx \zeta f_{\rm coll}(\mathbf{x})\left(1+\delta_{\rm L}(\mathbf{x})\right)  \notag\\
&\approx \zeta f_{\rm coll}(1+b_{\rm halo}\delta_{\rm L}(\mathbf{x}))
\end{align}
where $b_{\rm halo} = 1 + b_{\rm halo}^{\rm L}$ is the Eulerian halo bias, written in terms of the Lagrangian bias $b_{\rm halo}^{\rm L}$ which follows from the halo mass function \citep{1999MNRAS.308..119S}. Since photons are also conserved in this limit, we have $Q_{\rm HII}^M=\zeta f_{\rm coll}$. Thus $b_{\rm HII}$ in this limit is equal to the large-scale halo bias $b_{\rm halo}$ (and thus is larger for larger $M_{\rm min}$). Since the halo field is converged between different resolutions, so is $b_{\rm HII}$. 

There are also some interesting scalings apparent in the behaviour of $b_{\rm HII}$ as a function of $M_{\rm min}$ and resolution, which are essentially driven by the interplay between the physical sizes of the bubbles and the size of the grid cells, i.e., by the resolved fraction of bubbles or the fraction of partially ionized cells. E.g., the departure of $b_{\rm HII}$ from $b_{\rm halo}$, and thus the departure from convergence, occurs later for smaller $M_{\rm min}$ because the bubbles are smaller. Note that $b_{\rm HI} \to 0$ as $Q_{\rm HII}^M \to 1 / b_{\rm HII} = 1 / b_{\rm halo}$, at which point the high density regions are ionized in such a way that, when smoothed over large scales, the HI field shows no fluctuations. In simulations, however, $b_{\rm HI}$ does not become numerically equal to zero but only approaches a small numerical value.

As the bubbles begin to get resolved, the HII bias is given by a weighted average of $b_{\rm halo}$ and the bias $b_{\rm bub, res}$ of resolved bubbles. For coarser resolution, only larger bubbles contribute to $b_{\rm bub, res}$ and thus the bias is larger. At this stage, we find $b_{\rm HII}$ to decrease with increasing $\zeta$. This is easily understood in terms of the first crossing of the bubble barrier in the ES models. Since the bubble barrier height decreases with increasing $\zeta$ and it become ``easier'' for the walks to cross it thus making the bubbles less biased. The behaviour of the bias is different for larger $Q_{\rm HII}^M$ and $\zeta f_{\rm coll}$ as the non-linear effects begin to be dominant, hence the bias does not remain scale-free. In addition, the bubble sizes become comparable to the scale under consideration, hence the correlation is contributed by the so-called ``one-bubble term''.

\subsection{Photon non-conservation and bias non-convergence}
\label{app:photonNC->biasNC}

In the main text, we argued that photon non-conservation in the ES models implies that, at very high resolution and fixed ionizing efficiency $\zeta$, these models will produce a converged ionization field whose clustering properties will, in general, be different from the correct answer obtained at low resolution in the same models.

Here, we test this idea using high-resolution ionization maps. Since our approximate conditional mass-function approach does not work well at high resolution, for this exercise we use haloes obtained directly from the $N$-body simulation discussed in Appendix \ref{app:halofield}. This means that we are restricted to relatively large halo masses (determined by our mass resolution), and also relatively small length scales (determined by our box size). We therefore focus on scales $k\gtrsim 0.07 h/$cMpc and only generate maps with $M_{\rm min}=10^{10}\Msun$.

\fig{fig:unnormalized_bias_ionized_4panels_haloes} shows $(Q_{\rm HII}^Mb_{\rm HII})^2$ for maps produced using the ES model with different resolutions (solid curves) and with our PC model at two resolutions (dashed and dotted curves). The left and right columns show results for two values of $\zeta f_{\rm coll}$, while the top (bottom) row shows results for the ES model with the spherical tophat (sharp-$k$) filter. (The results of our PC model are identical across the two rows, and are clearly converged at large scales.) 

For the spherical tophat results, we see a clear validation of the arguments presented in \secn{subsec:photonNC->biasNC}. At small $\zeta f_{\rm coll}$ (top left panel), the low-resolution ES results converge to one value, while the high-resolution ES results also converge, but to a different value. As we showed above, the low-resolution ES answer is correct by construction and is determined entirely by the underlying halo bias. Our PC model has also clearly converged at large scales to the low-resolution ES value, which in fact serves as a validation of our PC algorithm. 

At higher $\zeta f_{\rm coll}$ (top right panel), bubble sizes are larger and we see that the $\Delta x=5h^{-1}$cMpc results, which were earlier ``low-resolution'', now behave like the high-resolution maps since more bubbles are resolved by this grid. We again see that the lowest-resolution ES result (which is the least photon non-conserving) is close to the (converged) PC results at large scales. The labels of the various ES curves also clearly show the increasing level of photon non-conservation as the resolution increases (i.e., as $\Delta x$ decreases). Thus, the results at both low and high $\zeta f_{\rm coll}$ are consistent with the arguments of \secn{subsec:photonNC->biasNC} tying the bias non-convergence of ES models directly to photon non-conservation\footnote{For much larger values of $\zeta f_{\rm coll}$, we find that the percolation of ionized regions in the simulation box becomes significant and hence the simple explanation based on bubble bias does not work any more. However, we have explicitly checked and found that the PC model even in this case produces a bias that is converged at large scales.}. 

The results for the sharp-$k$ filter in the bottom row of the Figure are more complicated, which is perhaps not very surprising since the filter is not localized in real space. This may lead to large-scale correlations in the ionization maps and as a result the intuition of spherical bubbles underlying the arguments of \secn{subsec:photonNC->biasNC} does not work for the sharp-$k$ filter. In particular, we find that both $\zeta f_{\rm coll} / Q_{\rm HII}^M$ and the large-scale bias exhibit non-monotonic behaviour with respect to the resolution, thus making the interpretation of the results less straightforward. Nevertheless, it is clear that the sharp-$k$ ES results at high resolution also have a high level of photon non-conservation and a bias that does not converge to the low-resolution value. 

\begin{figure}
\includegraphics[width=0.4\textwidth]{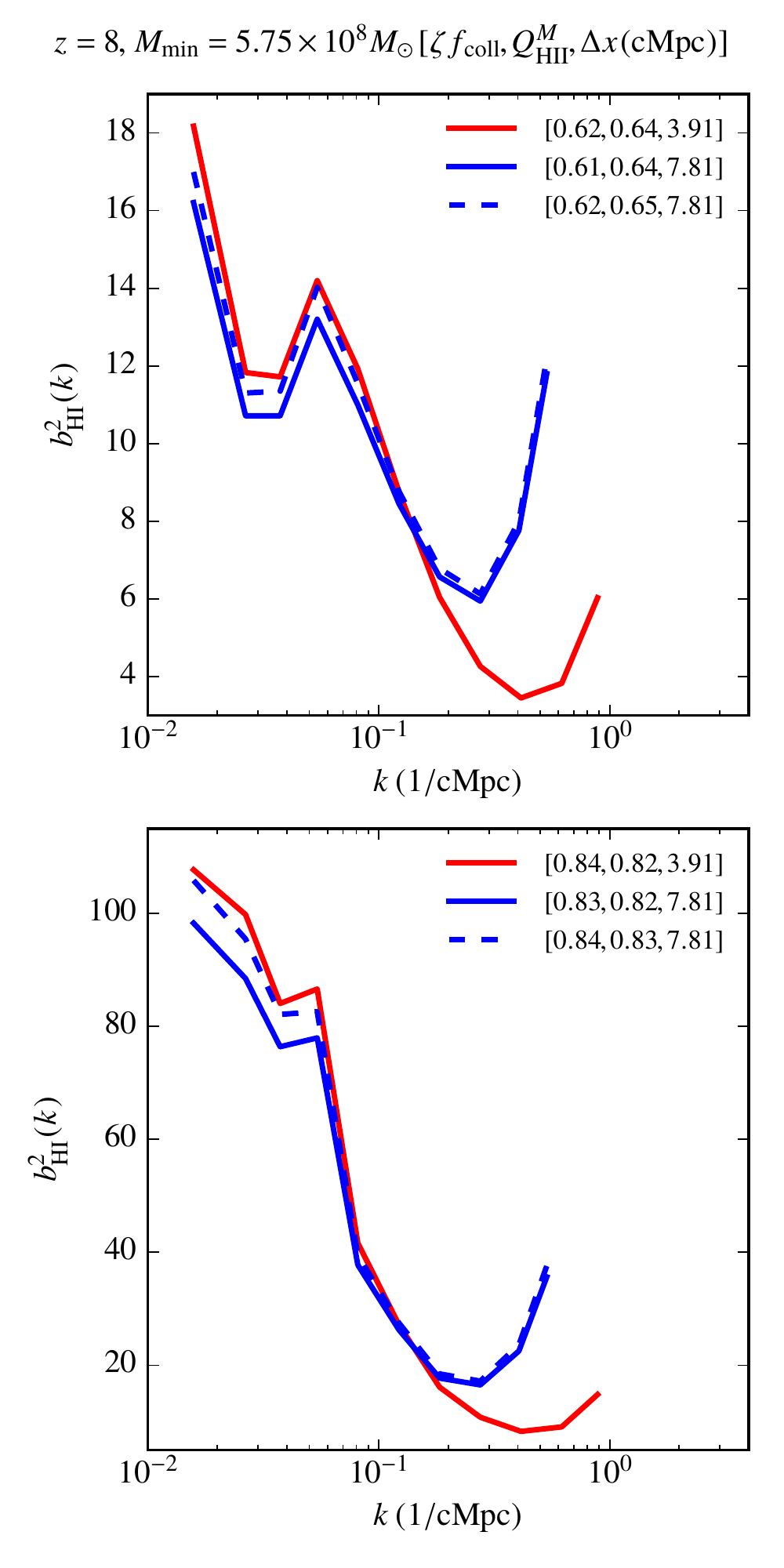}
\caption{
The bias of the HI density field $b_{\rm HI}^2(k)$ obtained from the 21cmFAST at $z = 8$ for different parameter values as indicated in the legend. 
}
\label{fig:bias_21cmFAST}
\end{figure}

We therefore conclude that our arguments in \secn{subsec:photonNC->biasNC} are indeed a correct description of the behaviour of ES models (at least until the ionized regions percolate significantly in the simulation volume).

\section{Large-scale bias in 21cmFAST}
\label{app:bias_21cmFAST}

In this Appendix, we examine whether the result obtained from our implementation of the ES-based semi-numerical model holds true for other such implementations as well. We carry out this exercise using the publicly available code 21cmFAST \citep{2011MNRAS.411..955M}. The fact that 21~cm maps obtained using 21cmFAST do not conserve photons was already pointed out by us in an earlier work \citep[see Appendix B of][]{2016MNRAS.460.1801P}. Our main aim here is to study the large-scale bias obtained using the 21cmFAST and check whether it converges between different map resolutions. We ensure that the changes made in the default parameters are minimum so as to avoid introducing any errors. The list of changes we make are as follows:

\begin{itemize}

\item We choose the cosmological parameters same as the rest of the paper. We take a box size $500$cMpc. The number of cells for sampling the initial conditions along a principal axis (i.e., the parameter \verb|DIM| in file \verb|Parameter_files/INIT_PARAMS.H|) is taken to be $768$.

\item We set the parameter \verb|INHOMO_RECO| in file \verb|Parameter_files/ANAL_PARAMS.H| to $0$ so as to disable any computation of inhomogeneous recombinations. We also set \verb|USE_TS_IN_21CM| in file \verb|Parameter_files/HEAT_PARAMS.H| to $0$ which is equivalent to assuming the spin temperature is much larger than the radiation temperature.

\item To generate maps of different resolutions, we set the parameter \verb|HII_DIM| in file \verb|Parameter_files/INIT_PARAMS.H| to the appropriate value. In this work, we have taken two values, namely $64$ (which is similar to our default resolution) and $32$ (the coarser resolution).

\end{itemize}

The results for $b_{\rm HI}^2(k)$ for two stages of reionization are shown in \fig{fig:bias_21cmFAST}. The values of different parameters can be read off from the legend. The amount of photon non-conservation can be found by looking at the values of $\zeta f_{\rm coll}$ and $Q_{\rm HII}^M$ for different curves. As one can see that $\zeta f_{\rm coll} / Q_{\rm HII}^M < 1$ when $Q_{\rm HII}^M = 0.5$, while the trend reverses to $\zeta f_{\rm coll} / Q_{\rm HII}^M > 1$ at later stages $Q_{\rm HII}^M = 0.8$. The amount of non-conservation is thus $\lesssim 5\%$, similar to our results for the sharp-$k$ filter. The large-scale bias for the coarse resolution map, however, does not converge to that for the high resolution one for $Q_{\rm HII}^M = 0.5$, the difference being $\sim 15-20\%$ ($\sim 5-10\%$) when the maps are normalized to the same value of $Q_{\rm HII}^M$ ($\zeta f_{\rm coll}$). The convergence is much better for $Q_{\rm HII}^M = 0.8$, consistent with what is found at late stages of reionization in our implementation of the ES-based algorithm.

We end this discussion by listing the differences between our implementation and that of 21cmFAST, which are as follows: (i) The effect of line-of-sight peculiar velocities are accounted for in 21cmFAST but not in our calculations, however, this is unlikely to make any differences to the relative bias at large-scales. (ii) There is a maximum horizon for the bubble size implemented in the 21cmFAST (taken to be $50$cMpc) which is not present in our method. This too is unlikely to make much difference to the maps since the typical bubble sizes are smaller than this horizon. (iii) The calculation of the collapsed fraction is implemented very differently in 21cmFAST than ours. It is calculated using the \citet{1991ApJ...379..440B} conditional mass function \citep[then scaled to match the mean $f_{\rm coll}$ of][]{1999MNRAS.308..119S}, while we use the conditional mass function of \citet{2002MNRAS.329...61S} (with parameter values re-adjusted to match simulations). Also, the collapse fraction calculation in 21cmFAST is done independently for each resolution, while we calculate it using the best resolution we decide to work on and then smooth the field for other coarser resolutions. In spite of these differences, it is clear that both the non-conservation of photons and non-convergence of large-scale bias are generic features of the ES-based semi-numerical models.

\label{lastpage}

\end{document}